# The role of homology in fluid vortices I: non-relativistic flow

D. H. Delphenich[†]
Spring Valley, OH 45370

**Abstract**: The methods of singular and de Rham homology and cohomology are reviewed to the extent that they are applicable to the structure and motion of vortices. In particular, they are first applied to the concept of integral invariants. After a brief review of the elements of fluid mechanics, when expressed in the language of exterior differential forms and homology theory, the basic laws of vortex theory are shown to be statements that are rooted in the homology theory of integral invariants.

([†]) E-mail: feedback@neo-classical-physics.info

# Contents

**Introduction.** – There has been an increasing interest in the subject of topological defects in ordered media in recent decades, as well as the associated concept of topology-changing processes. The applications of that mathematical technique have had two primary characters: applications to ordered media of the kind that can be readily addressed in laboratory experiments and engineering testing, which fall within the scope of condensed-matter physics, and applications to the kind of ordered media that have a more abstract definition, such as the space-time manifold in general, and often address speculative phenomena that might only take place in realms, such as the Planck scale, that might never be directly amenable to experimental tests.

To date, most of the literature on topological defects has been concerned with applying the methods of homotopy theory to the ordered media of condensed matter physics, except for some attempts to apply those techniques by analogy to general-relativistic and Planck-scale concepts. The main objectives of this monograph are to exhibit an example of how the methods of homology theory are more fundamental to the theory of topological defects and topology-changing processes than homotopy, and that the link is by way of the physical notion of conservation laws, balance principles, and integral invariants. The author made a similar attempt to establish the role of homology in the context of defective crystal lattices in a previous paper [**1**].

In particular, the applications of homology that will be addressed are directed towards the theory of vortices in fluids. The first article in the series will be confined to non-relativistic fluid mechanics, while the second one will address the relativistic kind. As it will turn out, some aspects of fluid mechanics are actually simpler in the relativistic context.

In either domain of applications to fluid motion, one will see that the places in which one is implicitly considering homological notions are in questions that pertain to the degree of integrability of flows and the aforementioned context of fundamental first principles such as the balance principle upon which the laws of motion are founded. This gives rise to absolute and relative integral invariants of the flow of fluid along the velocity and vorticity vector fields, and one finds that such fundamental laws as Helmholtz's and Kelvin's circulation theorems then take on a distinctly homological character in that way.

The first part of this article briefly reviews the basic definitions, constructions, and theorem of homology theory in both its singular form and its de Rham form, which is closest to the applications of fluid mechanics. In order to keep the volume of introductory material to a minimum, the judgment was made to assume that the reader had a minimal familiarity with abstract algebra beyond the usual group theory that is assumed of physicists. In particular, the concepts of rings and modules (which are essentially vector spaces whose scalars from rings that are more general than fields) are fundamental to the definitions of homology. Readers that are not familiar with those concepts might first confer some undergraduate text on abstract algebra, such as Birkhoff and MacLane [**2**] or Herstein [**3**]. Admittedly, it has been the author's experience that the making the jump from basic group theory to rings and modules is probably the main obstruction to going from homotopy to homology for most physicists. Furthermore, as Mermin pointed out in an article on the application of homotopy theory to topological defects in ordered media [**4**], even that level of applications involves a considerable degree of distillation of the physically-relevant facts from the abstractions and generalities of algebraic topology, followed by a process of tailoring them to the specialized demands of the problem at hand.

That is why we have attempted to keep to the aspects of homology and cohomology that are closest to the de Rham homology and cohomology that are an outgrowth of the calculus of exterior differential forms on differentiable manifolds. Hence, the basic notions of that calculus will have to be assumed to be familiar to the reader. In particular, since the fact that there even is such a thing as de Rham homology (but only for orientable manifolds), and that it is directly relevant to the first principles of physics, is not widely known, the author will have to introduce its basic notions, as well. Some of the relevant facts of de Rham cohomology can be found in other standard texts, such as de Rham [**5**], Warner [**6**], Goldberg [**7**], Vaisman [**8**], or Bott and Tu [**9**], although one will generally find that the definitions often tend to be made in a more general context than singular homology, which is closest to the process of integrating differential forms. Some standard texts on singular homology theory that one might benefit from are Hilton and Wylie [**10**], Vick [**11**], Greenberg [**12**], or Rotman [**13**], but, as mentioned above, there is a lot of distillation involved with reading them as a physicist, and to some extent that is one of contributions that this monograph hopes to achieve. Some other examples of the application of algebraic topology to engineering and physics can be found in Lefschetz [**14**] or Giblin [**15**].

Part I concludes with a discussion of how homology relates to the basic concept of time evolution, especially when one includes the possibility of topology-changing processes. Such processes occur regularly in fluid mechanics, such as the nucleation or cavitation of bubbles, the formation of smoke rings, and the formation of vortex pairs, so it is the author's steadfast belief that when one considers the possibilities that topology offers for wildly-pathological speculations on unphysical phenomena, one must agree that it would be best to establish how things work in a familiar natural context that is directly amenable to experiment before one reasons by analogy with phenomena that lie beyond direct observation. We have specialized (and generalized) the somewhat-more-familiar physical notion of Lorentz cobordism [**16**, **17**] to that of vector homology. The specialization is from cobordism, which is a generalized homology, to singular homology, which is more directly applicable to fluid mechanics, and the generalization is from the four dimensions of spacetime to arbitrary dimensions, as well as from non-singular vector fields to singular ones.

The second part of the present article begins by going over the basic definitions and laws of fluid mechanics when they are phrased in the language of exterior differential forms in order to make the application of homology more immediate. Some of the basic definitions are also covered in Arnol'd and Khesin [**18**], although their topological applications are of a fundamentally different character than the ones that are found here. Some of the standard texts on fluid mechanics that can prove useful here are Lamb [**19**], Batchelor [**20**], Milne-Thomson [**21**], and Landau and Lifschitz [**22**].

Finally, all of what was introduced in the main body of the text here is assembled into an explanation of how the basic definitions and laws of the non-relativistic theory of vortices can be explained in terms of the homology of motion. In particular, the homology of integral invariants is fundamental to the laws of circulation, which is basically a differentiable singular 1-cochain.

Part I: Mathematical preliminaries

**1. Elementary homology and cohomology.** – In this section, we shall introduce the basic constructions and theorems of singular homology and its dual cohomology, and then show how it relates to the de Rham homology and cohomology that one gets from the exterior differential forms on a differentiable manifold.

*a. Singular homology* [**10-13**]. The basic building blocks of singular homology are *singular k-simplexes*. Such an object is defined by a continuous map $\sigma_k : I^k \to M$, where $I = [0, 1]$, and $I^k$ is a reference $k$-cube and $M$ is a topological space.

Actually, we should be calling these objects singular *cubic* simplexes, since other reference objects are used, such as generalized tetrahedra and closed balls, which are all nonetheless homeomorphic to a $k$-cube. The fact that we have chosen the $k$-cube as a reference is based in the fact that one often encounters them in the context of the integration of exterior differential forms on manifolds, and we shall be dealing with those problems primarily.

The inclusion of the word "singular" is rooted in the fact that we have required only continuity of the map, not homeomorphism. Hence, there is nothing to stop the image of a singular $k$-simplex from degenerating to a lower-dimensional object than a $k$-cube, and possibly just a point. In the event that the map is a homeomorphism, one refers to such a "non-singular" $k$-simplex as a *k-cell*. One will ultimately find that when one passes to the homology modules, the contributions from the degenerate $k$-simplexes will cancel out and leave only the contributions from the $k$-cells.

The next things that one defines are *k-chains with coefficients in a ring R*, which are formal sums:

$$c_k = \sum_{i=1}^{N} a^i \sigma_k(i) \tag{1.1}$$

of $k$-simplexes $\sigma_k(i)$ with coefficients $a^i$ in $R$.

Rather than go into the abstract definition of a formal sum, we shall only specify some rules for calculating with them:

i) $\sum_{i=1}^{N} a^i \sigma_k(i) = \sum_{i=1}^{N} b^i \sigma_k(i)$ iff $a^i = b^i$ for all $i$.

ii) $\sum_{i=1}^{N} a^i \sigma_k(i) + \sum_{i=1}^{N} b^i \sigma_k(i) = \sum_{i=1}^{N} (a^i + b^i) \sigma_k(i)$.

iii) $\lambda \sum_{i=1}^{N} a^i \sigma_k(i) = \sum_{i=1}^{N} \lambda a^i \sigma_k(i)$.

iv) (1) $\sigma_k(i) = \sigma_k(i)$,

v) (0) $\sigma_k(i) =$ "0" (i.e., the term no longer appears in the summation).

Because of the last property, one can extend *i*) and *ii*) to the case in which the sets of basic *k*-simplexes in the two sums are not identical by introducing 0 coefficients into each sum for the simplexes that are not common to both of them.

One sees that the basic simplexes are playing a role in these sums that is analogous to that of basis vectors in a vector space. Indeed, in the case where $R$ is the field $\mathbb{R}$ of real numbers, they will be scalar combinations of vectors. More generally, the set of all singular *k*-simplexes (which is bound to be uncountably infinite, in most cases) defines the generators for the *free R-module* $C_k(M; R)$ of all singular *k*-chains in $M$ with coefficients in $R$; when $R = \mathbb{R}$, it will be a free vector space over that set of basis vectors.

Any singular cubic *k*-simplex $\sigma_k$ will have a boundary, which will be a $(k-1)$-chain:

$$\partial \sigma_k = \sum_{a=1}^{2k} \pm \sigma_{k-1}(a) \tag{1.2}$$

with coefficients that equal ± 1, and as many terms as there are faces to the *k*-cube, namely, 2*k*. The signs come from the fact that we are "orienting" $\partial \sigma_k$ in such a way that the $a^{th}$ *pair of faces* of $I^k$, namely, (…, 0, …) and (…, 1, …), where the 0 and 1 appear in the $a^{th}$ coordinate, are given the – and + signs respectively. Hence, for a 1-simplex $\sigma_1$: [0, 1] $\to M$, the boundary will be $\partial \sigma_1 = \sigma_0(1) - \sigma_0(0)$, for a 2-simplex $\sigma_2$ : [0, 1] × [0, 1] $\to$ $M$ the boundary will be $\partial \sigma_2 = \sigma_1(1, s) - \sigma_1(0, s) + \sigma_1(s, 1) - \sigma_1(s, 1)$, etc.

Once one has defined the boundary of each basic simplex, one can define the boundary of any *k*-chain by "extending by linearity"; i.e. if $c_k$ is as in (1.1):

$$\partial c_k = \sum_{i=1}^{N} a^i \partial \sigma_k(i) = \sum_{a=1}^{N_{k-1}} \left( \sum_{i=1}^{N_k} a^i \partial_i^a \right) \sigma_{k-1}(a). \tag{1.3}$$

In this, we have defined an $N_k \times N_{k-1}$ matrix $\partial_i^a$ whose elements are – 1, 0, or + 1, and which can be called the *incidence matrix* of $\partial c_k$ with respect to the set of basic $(k-1)$-simplexes $\sigma_{k-1}(a)$. Hence, if $b^a$ is the coefficient of $\sigma_{k-1}(a)$ in $\partial c_k$ then:

$$b^a = \sum_{i=1}^{N_k} a^i \partial_i^a . \tag{1.4}$$

The expression (1.3) is deceptive in its simplicity, since the terms in (1.1) generally involve basic simplexes that have common faces, which will generally cancel when one takes the boundary of $c_k$. One then thinks of those boundary faces as having been "identified," and in fact, that is a common way of representing elementary manifolds with boundary as singular *k*-chains.

For example, one can represent the surface of a two-dimensional cylinder naively as the images of the vertical faces of a 3-cube that connect the $z = 0$ face to the $z = 1$ face, but only when one understands that when one takes the boundary of that 2-chain, the common edges will cancel. That is, let the cylinder be represented by the 2-chain:

$$c_2 = \sigma_2(1, y, z) + \sigma_2(0, y, z) + \sigma_2(x, 1, z) + \sigma_2(x, 0, z),$$

which represent the four vertical faces of the unit 3-cube. We illustrate this situation in Fig. 1.

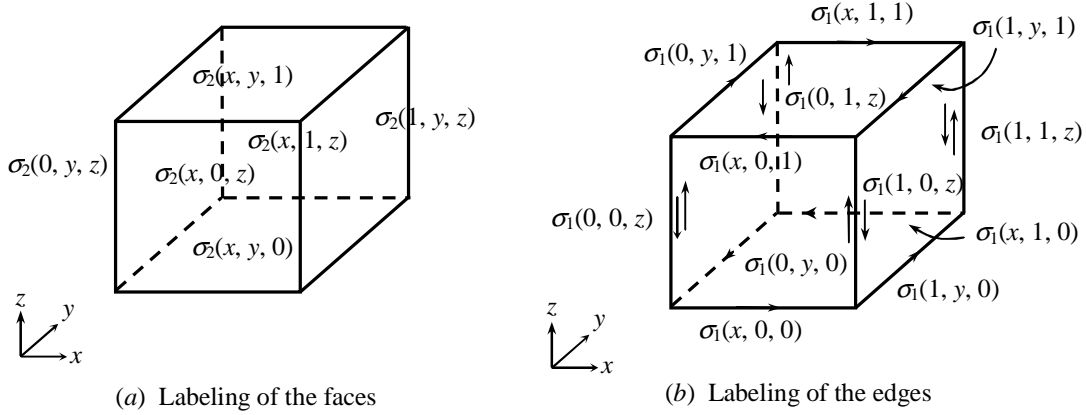

(a) Labeling of the faces  (b) Labeling of the edges

Figure 1. Representing a 2-cylinder by the vertical faces of a 3-cube.

When one takes the boundary of this, one will get:

$$\begin{aligned}\partial c_2 = &\ \sigma_1(1, 1, z) - \sigma_1(1, 0, z) - \sigma_1(1, y, 1) + \sigma_1(1, y, 0) \\ &- \sigma_1(0, 1, z) + \sigma_1(0, 0, z) + \sigma_1(0, y, 1) - \sigma_1(0, y, 0) \\ &+ \sigma_1(x, 1, 1) - \sigma_1(x, 1, 0) - \sigma_1(1, 1, z) + \sigma_1(0, 1, z) \\ &+ \sigma_1(1, 0, z) - \sigma_1(0, 0, z) - \sigma_1(x, 0, 1) + \sigma_1(x, 0, 0) \\ \\ = &- \sigma_1(1, y, 1) + \sigma_1(x, 1, 1) - \sigma_1(x, 0, 1) + \sigma_1(0, y, 1) \\ &+ \sigma_1(1, y, 0) - \sigma_1(x, 1, 0) + \sigma_1(x, 0, 0) - \sigma_1(0, y, 0).\end{aligned}$$

This 1-chain then represents the two squares that are defined by the top and bottom boundaries of the vertical faces of the cube.

Since we are dealing with continuous maps, we can actually simplify the representation of the 2-cylinder to even fewer 2-simplexes by "bending" them and identifying the boundary points. For instance, we can represent a cylinder as a 2-chain:

$$c_2 = \sigma_2(1) + \sigma_2(2),$$

whose boundary will be:

$$\partial c_2 = \partial \sigma_2(1) + \partial \sigma_2(2) = \sigma_1(1) - \sigma_1(2) + \sigma_1(3) - \sigma_2(4)$$

if we also have:

$$\begin{aligned}\partial \sigma_2(1) = &\ \sigma_1(1) + \sigma_1(5) - \sigma_2(4) - \sigma_1(6), \\ \partial \sigma_2(2) = &- \sigma_1(2) + \sigma_1(6) - \sigma_1(3) - \sigma_1(5).\end{aligned}$$

We illustrate this process in Fig. 2:

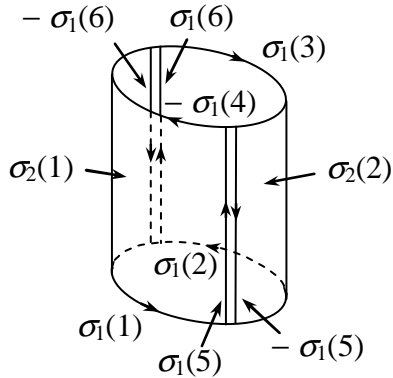

Figure 2. The representation of a cylinder by a formal sum of two 2-simplexes whose edges are identified by the boundary operator.

When one can represent a differentiable manifold $M$ as set of 0-chains, 1-chains, …, $n$-chains that are connected to each other by a suitably-defined boundary operator $\partial$, one says that $M$ has been *triangulated*, even when the basic simplexes are not triangular, tetrahedral, etc. According to Munkres [**23**], this is always possible for finite-dimensional manifolds (up to homotopy equivalence), even when $M$ is not compact. However, if $M$ is compact then it will always admit a *finite* triangulation (i.e., a finite number of chains in each dimension). A contractible space, such as a vector space, is homotopically equivalent to a point ([1]), which can be triangulated by a single 0-simplex, so that would give an example of a non-compact, differentiable manifold that admitted a finite triangulation. Another simple example that we shall use is that of a punctured plane $\mathbb{R}^2 - P$, which has any circle that surrounds the missing point $P$ as a deformation retraction, and that circle can be triangulated by two 1-simplexes whose end points are suitably identified by the boundary operator.

If we were to take the boundary of $\partial c_k$ in any of the examples above then we would find that it would vanish in every case. In fact, that is a general property of the boundary operator, namely, that:

$$\partial^2 = 0. \tag{1.5}$$

Hence, the boundary operator can be characterized by being a linear operator $\partial : C_k(M; R) \to C_{k-1}(M; R)$ for each $k$, with the property (1.5). One can also express this as a sequence of maps:

$$\ldots \to C_k(M; R) \xrightarrow{\partial} C_{k-1}(M; R) \xrightarrow{\partial} \ldots \to C_0(M; R) \xrightarrow{\partial} 0.$$

---

([1]) Recall (cf., e.g., Munkres [**24**]) that $f: M \to N$ is a *homotopy equivalence* iff there is a $g : N \to M$ such that $f \cdot g$ is homotopic to the identity on $M$ and $g \cdot f$ is homotopic to the identity on $N$. If only the first condition is satisfied then $f$ is a *deformation retraction* of $M$ to the subspace $g(N)$. A *contraction* of $M$ is a homotopy of the identity map to the inclusion of a point; if such a map exists then $M$ will be called *contractible*.

The condition (1.5) only implies that at each step in this sequence the image of the boundary map to the left (which will then be a sub-module of the module at that step) must be a sub-module of the kernel of the next boundary map (which will also be a sub-module). Whether or not it is *exact* – i.e., whether or not the image *equals* the kernel – will depend upon the *homology module* in that dimension, and we shall now clarify that notion.

The kernel of $\partial : C_k(M; R) \to C_{k-1}(M; R)$ is a submodule $Z_k(M; R)$ of $C_k(M; R)$ that consists of all $k$-chains $z_k$ that have vanishing boundaries, and are referred to as *k-cycles*. Hence:

$$\partial z_k = 0. \tag{1.6}$$

The image of $\partial : C_{k+1}(M; R) \to C_k(M; R)$ is a submodule $B_k(M; R)$ of $C_k(M; R)$ that consists of all $k$-chains $b_k$ that are *k-boundaries*. Hence:

$$b_k = \partial c_{k+1} \qquad [\text{for some } c_{k+1} \in C_{k+1}(M; R)]. \tag{1.7}$$

The $k+1$-chain $c_{k+1}$ is not generally unique, although any two candidates will have to "cobound," which we shall define shortly.

One can then say that every boundary is a cycle, but not always the converse. In order to characterize the difference between cycles and boundaries, we introduce two closely-related equivalences that one can define between $k$-chains $c_k$ and $c'_k$.

The first one is that they might be *cobounding*; i.e.:

$$\partial c_k = \partial c'_k. \tag{1.8}$$

One sees immediately that:

**Proposition:** $c_k$ and $c'_k$ *are cobounding iff* $c'_k - c_k$ *is a k-cycle.*

This sort of situation shows up frequently in the context of the work functional, where one might consider distinct paths – i.e., two 1-chains – between two points. The work that is done is then independent of the paths between those two points iff it is zero around the 1-cycle that they define when one traverses the first one of them directly and then the other one in reverse order.

A stronger condition to impose upon the chains $c_k$ and $c'_k$ is that they should be *homologous*; i.e., that $c'_k - c_k$ is a $k$-boundary. To continue with the work example, the paths between two points would be homologous iff they were homotopic.

Since the difference of two cycles is always a cycle, it is the last equivalence that will allow one to distinguish equivalence classes of cycles. One defines the *homology class* of a $k$-cycle $z_k$ to be the equivalence class $[z_k]$ of all $k$-cycles that differ from $z_k$ by a $k$-boundary; i.e., all $z'_k \in Z_k(M; R)$

$$z'_k - z_k = \partial c_{k+1} \tag{1.9}$$

for some $c_{k+1} \in C_{k+1}(M; R)$.

Naturally, the $k+1$-chain $c_{k+1}$ is not unique, since it, too, belongs to an equivalence class of cobounding $k+1$-chains.

The quotient module $H_k(M; R) = Z_k(M; R) / B_k(M; R)$, which then consists of all these equivalence classes $[z_k]$, is referred to as the *k-dimensional homology module* for $M$ with coefficients in $R$. In order to get some intuitive idea of what its elements represent, one can consider some low-dimensional examples of $k$-cycles that do not bound.

In dimension zero, one first finds that all 0-simplexes are closed. As it turns out, the generators of $H_0(M; R)$ are in one-to-one correspondence with the path components of $M$. Hence, $H_0(M; R) \cong R$ iff $M$ is path-connected.

One of the simplest examples in dimension one is that of a circle that does not bound a disc. One can find such a thing in the punctured plane, since there are two types of circles (i.e., 1-cycles) in the punctured plane: ones that do not surround the missing point and ones that do. The former circles bound discs, and are therefore homologous to 0, while the latter circles do not bound discs, but are always homologous to each other, and thus define the non-vanishing generator of $H_1(\mathbb{R}^2 - P; R) \cong H_1(S^1; R) \cong R$.

In dimension two, one looks for examples of 2-cycles (such as 2-spheres) that do not bound 3-chains (such as closed 3-balls). This time, one can consider singly-punctured $\mathbb{R}^3$, which has two types of spheres: ones that do not surround the missing point and ones that do. This becomes a higher-dimensional analogue of the previous one-dimensional case, and in fact, the two-dimensional homology module of punctured $\mathbb{R}^3$ has one generator that is defined by all of the spheres that surround the missing point.

This gives one the general impression that generators for $H_k(M; R)$ will represent "$k$-dimensional holes" in $M$, in a manner of speaking. Actually, there are two types of holes: holes with torsion and holes with no torsion. This corresponds to the fact that although both $Z_k(M; R)$ and $B_k(M; R)$ are free modules, their quotient module does not have to be free, and will generally split into a free submodule and one that has torsion elements. We shall not, however, elaborate on the latter possibility, since the only ring $R$ that we shall be concerned with is the field $\mathbb{R}$, and since the modules in question will all be vector spaces, $H_k(M; \mathbb{R})$ will have no torsion generators.

Furthermore, another useful property of $H_k(M; R)$ is that it is often finitely-generated, even though both $Z_k(M; R)$ and $B_k(M; R)$ will generally have an uncountable infinitude of generators; indeed, $H_k(M; R)$ can often vanish outright. In particular, if $M$ is compact then $H_k(M; R)$ must be finitely-generated in every dimension, but punctured $\mathbb{R}^n$ gives a non-compact example of finitely-generated homology. When $R = \mathbb{R}$, $H_k(M; \mathbb{R})$ can even be a finite-dimensional vector space.

An important property of singular homology is that since a singular $k$-simplex is simply a continuous map $\sigma_k : I^k \to M$, if one has a continuous map $f : M \to N$ then the composition $f \cdot \sigma_k : I^k \to N$ will be a $k$-simplex in $N$. Hence, there is a "push-forward" map $f_* : C_*(M; R) \to C_*(N; R)$ that one defines by extending the maps $f \cdot \sigma_k$ by linearity.

Moreover, if $f$ and $g$ are two homotopic maps from $M$ to $N$ then those push-forward maps will induce the same linear map in homology: $[f_*], [g_*] : H_*(M; R) \to H_*(N; R)$; i.e.:

**Theorem:** *If f, g: $M \to N$ are homotopic then $[f_*] = [g_*]$.*

In particular:

**Theorem:** *If $f : M \to N$ is a deformation retraction of $N \subset M$ then $H_*(M; R) \cong H_*(N; R)$.*

**Corollary**: *If $f : M \to N$ is a homotopy equivalence then $H_*(M; R) \cong H_*(N; R)$.*

For instance, any contractible space, such as any vector space, will have the same homology modules as a space that consists of a single point, and all of those modules vanish, except for $H_0(M; R) \cong R$, which has one generator.

The converse of the last theorem is not always true, and in that sense, homology is somewhat more "coarse-grained" than homotopy as an equivalence. That is, things can be homologous without being homotopic.

Of particular interest to the theory of vortices is the example of $M = \mathbb{R}^3$ minus a line, such as the $z$-axis. One can first make a deformation retraction of $M$ to $\mathbb{R}^2$ minus a point without changing its homology. One can then make another deformation retraction of the punctured plane to a circle $S^1$ that surrounds the point, and still have the same homology modules; i.e., $H_k(S^1; R)$, which will vanish except in dimensions zero and one, where one will have one generator.

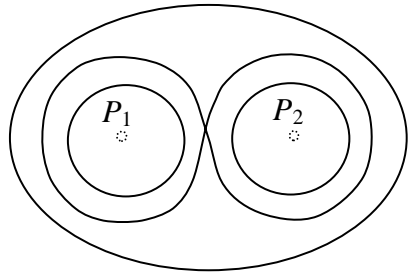

Figure 3.

The case of more than one vortex line is slightly more involved. Basically, one observes that a continuous loop is also a singular 1-cycle. If two loops are homotopic then the homotopy will define a singular 2-chain whose boundary consists of those two 1-cycles; i.e., they will be homologous. Hence, there is a map from $\pi_1(M)$ to $H_1(M;R)$ that takes each homotopy class of loops to a homology class of 1-cycles. However, there is a significanct difference between the group structure of $\pi_1(M)$, which can be non-Abelian, and the underlying additive group of $H_1(M;R)$, which is always Abelian. One finds that that the underlying additive group of $H_1(M;R)$ is isomorphic to the Abelianization of $\pi_1(M)$, which is the quotient of $\pi_1(M)$ by its commutator subgroup. We illustrate some of the possibilities for non-homotopic loops in a doubly-punctured plane in Fig. 3.

One sees that the loops that encircle each point by itself will clearly define homotopy classes in $\pi_1(M)$, and thus generators of $H_1(M;R)$. However, under Abelianization, the figure-eight loop that encircles both points will become a formal sum of two basic 1-cycles, as well as the outer loop. Hence, $H_1(M;R)$ will have one generator for each point.

A common construction that one encounters in the theory of vortices is the introduction of "cuts" into a multiply-connected space in order to render it simply-connected. In the previous example of a doubly-punctured plane, we expand the points $P_1$ and $P_2$ to finite circles, and see that if we introduce cuts as in Fig. 4 then the remaining space will be simply-connected. Since each cut effectively breaks any generating 1-cycle, and the 1-cycles that it cuts are all homologous, there is then a one-to-one correspondence between the cuts and the generators of $H_1(M;R)$. Hence, one can think of the cuts that one encounters in many treatments of vortices (e.g., [**19-22, 25-28**]) as being another way of representing the generators of $H_1(M;R)$.

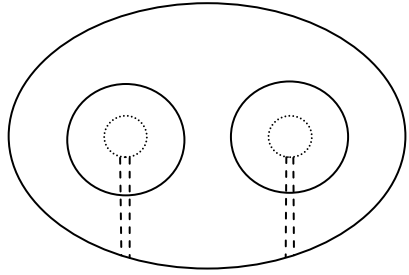

Figure 4.

*b. Singular cohomology*. Cohomology is dual to homology in the same way that the vector space $V^*$ of linear functionals on a vector space $V$ is dual to $V$. That is, one starts by defining the $R$-module $C^k(M; R)$ of homomorphisms of $C_k(M; R)$ into $R$ (i.e., linear functionals); its elements $c^k$ are referred to as *k-cochains with values in R*. Just as $C_k(M; R)$ is generated by singular $k$-simplexes $\sigma_k$, $C^k(M; R)$ is generated by dual $k$-simplexes $\sigma^k$, which are linear functionals on $k$-chains that have the property that for each $k$-simplex $\sigma_k(\alpha)$ there is a unique dual $k$-simplex $\sigma^k(\alpha)$ such that when one evaluates $\sigma^k(\alpha)$ on any $k$-simplex $\sigma_k(\beta)$, one will get:

$$<\sigma^k(\alpha), \sigma_k(\beta)> = \delta_{\alpha\beta}, \qquad (1.10)$$

in which the bracket notation is introduced for the bilinear pairing $C^k(M; R) \times C_k(M; R) \to R$ that is defined by the evaluation of functionals on elements.

Thus, a $k$-chain will take the form of a finite formal sum with coefficients in $R$:

$$c^k = \sum_{i=1}^{N} a_i \sigma^k(i). \qquad (1.11)$$

If we add zero coefficients in order to make a $k$-chain $c_k = \sum_{i=1}^{N} b^i \sigma_k(i)$ have basic simplexes that are dual to those of $c^k$ then the evaluation of $c^k$ on $c_k$ will be:

$$<c^k, c_k> = \sum_{i=1}^{N} a_i b^i. \qquad (1.12)$$

One defines a *coboundary* operator on $k$-cochains that is adjoint to the boundary operator on $k$-chains under the bilinear pairing. Namely, for each $k$, one has an $R$-linear map $\delta: C^k(M; R) \to C^{k+1}(M; R)$ that is defined by the condition that:

$$<\delta c^k, c_{k+1}> = <c^k, \partial c_{k+1}> \tag{1.13}$$

in any case.

One finds that $\delta$ has a property that is dual to that of $\partial$, namely:

$$\delta^2 = 0. \tag{1.14}$$

[The proof of that statement is by repeated use of (1.13).]

Hence, one can define $R$-modules using the images and kernels of $\delta$ that are analogues of the modules of cycles, boundaries and homology classes: A $k$-cochain $c^k$ is a *k-cocycle* iff $\delta c^k = 0$, so the kernel of $\delta: C^k(M; R) \to C^{k+1}(M; R)$ is an $R$-module $Z^k(M; R)$ that consists of all $k$-cocycles. $c^k$ is a *coboundary* iff there is some $k-1$-cochain $c^{k-1}$ such that $\delta c^{k-1} = c^k$, and the image of $\delta: C^{k-1}(M; R) \to C^k(M; R)$ is an $R$-module $B^k(M; R)$ that consists of all $k$-coboundaries. The quotient $R$-module $H^k(M; R) = Z^k(M; R) / B^k(M; R)$ is then called the *k-dimensional cohomology module of M with values in R*. Its elements $[z^k]$ consist of equivalence classes of $k$-cycles that differ by some $k$-coboundary:

$$\bar{z}^k - z^k = \delta c^{k-1}. \tag{1.15}$$

As with homology, the $k-1$-cochain $c^{k-1}$ will not be unique, and other possibilities for it will differ by a $k-1$-cocycle:

$$\bar{c}^{k-1} - c^{k-1} = z^{k-1} \qquad (\text{so } \delta \bar{c}^{k-1} = \delta c^{k-1}). \tag{1.16}$$

Just as the homology modules $H_k(M; R)$ generally decomposed into a free $R$-module and one that consisted of torsion elements, the same can be said of $H^k(M; R)$. In fact, it is only if there are no torsion elements in $H^k(M; R)$ that one can say that it is truly the dual $R$-module $\text{Hom}(H_k(M; R); R)$ of $R$-linear functionals on $k$-chains. Fortunately for us, most of our applications will involve the case in which $R = \mathbb{R}$, and just as for homology, that will make $H^k(M; \mathbb{R})$ into the real vector space is dual to $H_k(M; \mathbb{R})$. In particular, when $H_k(M; \mathbb{R})$ is a finite-dimensional vector space (such as for compact $M$), $H^k(M; \mathbb{R})$ will be linearly isomorphic to it, although the isomorphism will not be unique. However, if one is given a specific finite basis for $H_k(M; \mathbb{R})$, one can define the reciprocal basis on $H^k(M; \mathbb{R})$, and the linear isomorphism that the association of the one basis with the other one represents will take the coefficients $a^i$ of a $k$-chain $\sum a^i \sigma_k(i)$ to the coefficients $a_i$ of the $k$-cochain $\sum a_i \sigma^k(i)$ in such a way that $a^i = a_i$; i.e., one has basically transposed the column vector whose components are $a^i$ into the row vector whose components are $a_i$.

When *M* is an orientable, *n*-dimensional manifold, there is a type of duality between homology and cohomology in the form of isomorphisms between $H_k(M; R)$ and $H^{n-k}(M; R)$ that one calls *Poincaré duality*. However, since it basically originates in the volume element, we shall find that it is simpler to explain in the context of de Rham homology and cohomology, and we now pass on to those topics.

*c. De Rham cohomology* [**5-9**]. Although we introduced singular homology first above, and then introduced singular cohomology as its dual, since most mathematical texts and articles introduce only de Rham cohomology as something that is naturally defined by the calculus of exterior differential forms on a differentiable manifold and never introduce the homology that is dual to it (at least, for orientable manifolds), we shall follow that sequence, while also defining de Rham homology.

The *k-cochains* of de Rham cohomology for an *n*-dimensional differentiable manifold *M* are the exterior *k*-forms on it, which then define the infinite-dimensional real vector space $\Lambda^k(M)$, which plays the role of $C^k(M; \mathbb{R})$ in singular homology. Since the exterior derivative operator is a linear map $d: \Lambda^k(M) \to \Lambda^{k+1}(M)$ with the property that $d^2 = 0$, one sees that it certainly behaves like a coboundary operator. This defines the *Poincaré sequence*:

$$0 \to \Lambda^n(M) \xrightarrow{d} \ldots \xrightarrow{d} \Lambda^k(M) \xrightarrow{d} \Lambda^{k+1}(M) \xrightarrow{d} \ldots \xrightarrow{d} \Lambda^0(M) \to 0$$

Once again, image of *d* in $\Lambda^{k+1}(M)$ is a subspace of the kernel of the next *d* operator, but it is not necessarily equal to that subspace.

A *k*-form $\alpha$ is called *closed* iff $d\alpha = 0$ and *exact* iff there is a *k*−1-form $\beta$ such that $\alpha = d\beta$. Of course, $\beta$ is not unique, and any other $\beta'$ such that $d\beta' = d\beta$ will differ from $\beta$ by a closed *k*−1-form $z^k$:

$$\beta' - \beta = z^k. \tag{1.17}$$

According to the *Poincaré lemma*, the fact that every point of a manifold has a contractible neighborhood (e.g., a coordinate neighborhood) implies that every closed differential form is locally exact, so the Poincaré sequence will be locally exact; i.e., when one restricts *M* to a contractible neighborhood.

The vector subspace of all closed *k*-forms in $\Lambda^k(M)$ (i.e., the kernel of *d*) will be denoted by $Z^k_{dR}(M)$ and the vector space of all exact ones (i.e., the image of *d*) will be denoted by $B^k_{dR}(M)$, and the quotient vector space $H^k_{dR}(M) = Z^k_{dR}(M)/B^k_{dR}(M)$ will be called the *de Rham cohomology vector space in dimension k*. Its elements consist of equivalence classes $[z^k]$ of closed forms that are cohomologous; i.e., they differ by exact forms:

$$\bar{z}^k - z^k = d\beta \quad \text{for some } k-1\text{-form } \beta. \tag{1.18}$$

One can also characterize a de Rham cohomology class by saying that if it is 0 then it consists of nothing but all exact *k*-forms, while if it is non-trivial, its elements will all be closed *k*-forms that are not exact.

Whereas $Z_{\text{dR}}^k(M)$ and $B_{\text{dR}}^k(M)$ are infinite-dimensional, $H_{\text{dR}}^k(M)$ can often be finite-dimensional, if not 0. Once again, if $M$ is compact then $H_{\text{dR}}^k(M)$ will always be finite-dimensional. In order to see what its generators (i.e., basis elements) consist of, we need to discuss:

**De Rham's theorem:** *For every k:*

$$H_{\text{dR}}^k(M) \cong H^k(M; \mathbb{R}). \qquad (1.19)$$

The basis for this isomorphism is in the fact that since the integration of $k$-forms is linear on formal sums of $k$-simplexes, the integration of a $k$-form $\alpha$ on a $k$-chain $c_k$ will define an $\mathbb{R}$-linear functional on $k$-chains with coefficients in $\mathbb{R}$:

$$\alpha[c_k] = \int_{c_k} \alpha. \qquad (1.20)$$

Hence, $\alpha[.]$ is a singular $k$-cochain with values in $\mathbb{R}$, and that will define a linear map $\Lambda^k(M) \to C^k(M; \mathbb{R})$, $\alpha \mapsto \alpha[.]$.

In order relate the de Rham coboundary operator (viz., $d$) to the singular one $\delta$, one can use Stokes's theorem. This time, let $\alpha$ be a $k-1$-form:

$$\delta\alpha[c_k] = \alpha[\partial c_k] = \int_{\partial c_k} \alpha = \int_{c_k} d\alpha = d\alpha[c_k].$$

Hence, $\delta$ does the same thing to singular $k$-cochains with values in $\mathbb{R}$ that $d$ does to $k$-forms. We now have linear maps $Z_{\text{dR}}^k(M) \to Z^k(M; \mathbb{R})$, $B_{\text{dR}}^k(M) \to B^k(M; \mathbb{R})$, which are restrictions of the map $\Lambda^k(M) \to C^k(M; \mathbb{R})$.

It is now easy to see why $H^k(M; \mathbb{R}) = 0$ for every $k > n$, since under the de Rham isomorphism, that will correspond to the fact that there are no non-zero exterior $k$-forms for $k > n$. That also implies that all $n$-forms will be closed; i.e., $Z_{\text{dR}}^n(M) = \Lambda^n(M)$. Whether they are also exact will depend upon the orientability of $M$, which we shall discuss below.

One can characterize de Rham cocycles and coboundaries by the way that they integrate over singular cycles and boundaries:

**Theorem:**

i) *A k-form $\alpha$ is a de Rham cocycle iff:*

$$\int_{b_k} \alpha = 0$$

*for every k–1-boundary $b_k$ .*

ii) *A k-form $\alpha$ is a de Rham coboundary iff:*

$$\int_{z_k} \alpha = 0$$

*for every k-cycle $z_k$ .*

Since de Rham cohomology is isomorphic to a cohomology with values in a field, it will have no torsion elements. Thus, de Rham cohomology is even more coarse-grained than singular cohomology with values in more general rings, such as the integers. Although this means that, in particular, the difference between an *n*-sphere and *n*-dimensional real projective space will not show up in de Rham cohomology, since it takes the form of a $\mathbb{Z}_2$ torsion summand in singular cohomology with integer values, nonetheless, one will also find that since the basic notions of de Rham cohomology were present in the mathematics of many topics in physics all along, the advantages will often outweigh the disadvantage that its topological ambiguities represent.

An example of a 1-form that is closed, but not exact, that will be useful in the study of vortices is given by the angle 1-form $\dbar\theta$ on the punctured plane. We denote with the thermodynamic notation of using $\dbar$, instead of $d$, to emphasize that it is not truly an exact differential.

The fact that it is closed follows from the fact that if $b_1$ bounds a 2-cycle then $\int_{b_1} \dbar\theta$ will equal the winding number of $b_1$ about the missing point *P*, which will be zero. We illustrate this in Fig. 5.

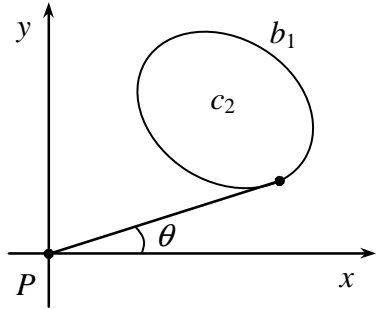

Figure 5. Integrating the angle 1-form around a bounding 1-cycle.

The fact that it is not exact follows from the fact that if *C* is a unit circle that surrounds *P* then

$$\int_C \dbar\theta = 2\pi \neq 0. \qquad (1.21)$$

This is dual to the fact that *C* does not bound a disc in this case, and thus [*C*] defines a generator of $H_1(M; \mathbb{R}) \cong \mathbb{R}$. One can introduce a cut into the plane that extends from the origin along the *x*-axis so that $\theta$ becomes single-valued on the complement of the cut and has a jump discontinuity across it. The cut then represents the basis element of $H_1(M; \mathbb{R})$, and since the linear isomorphisms $H_1(M; \mathbb{R}) \cong H^1(M; \mathbb{R}) \cong H^1_{dR}(M)$ are easy to define in the one-dimensional case, one sees that $d\theta$ is the basis element of $H^1_{dR}(M)$ that corresponds to either non-bounding circles in the punctured plane or cuts in it that render it simply-connected.

*d. De Rham homology.* When an *n*-dimensional differentiable manifold *M* is orientable, it will also admit a *volume element.* Such an object is simply a globally non-zero *n*-form *V*. Since all *n*-forms on an *n*-dimensional manifold must be closed, *V* will also define a de Rham cohomology class [*V*] in dimension *n*, and, in fact, $H^n_{dR}(M) \cong \mathbb{R}$ iff *M* is compact and orientable; otherwise, $H^n_{dR}(M) = 0$. That explains why a two-dimensional space such as the punctured plane can have the same homology as a one-dimensional space such as the circle: Although both are orientable spaces, the punctured plane is not compact, so $H^2_{dR}(\mathbb{R}^2 - P) = 0$.

If $(U, x^i, i = 1, \ldots, n)$ is a local coordinate system on *M* then one can represent *V* in terms of the natural coframe field $\{dx^i, i = 1, \ldots, n\}$ as:

$$V = dx^1 \wedge \ldots \wedge dx^n = \frac{1}{n!} \varepsilon_{i_1 \cdots i_n} dx^{i_1} \wedge \cdots \wedge dx^{i_n}, \qquad (1.22)$$

in which $\varepsilon_{12\ldots n}$ is the completely-symmetric Levi-Civita symbol.

The volume element *V*, in conjunction with the interior product of an exterior form by a multivector field, allows one to define linear isomorphisms $\# : \Lambda_k \to \Lambda^{n-k}$ for each *k*. Namely, if $\mathbf{A} \in \Lambda_k$ is a *k*-vector field on *M* then the *n−k*-form $\#\mathbf{A}$ will be defined by:

$$\#\mathbf{A} = i_\mathbf{A} V. \qquad (1.23)$$

Hence, for any *n−k*-vector field **B**, the evaluation of $\#\mathbf{A}$ on **B** will give:

$$\#\mathbf{A}(\mathbf{B}) = V(\mathbf{A} \wedge \mathbf{B}). \qquad (1.24)$$

One can call these isomorphisms the *Poincaré isomorphisms*, although that term is usually first introduced in the context of homology nowadays.

With the local form for *V* above in (1.22), if:

$$\mathbf{A} = \frac{1}{k!} A^{i_1 \cdots i_k} \partial_{i_1} \wedge \cdots \wedge \partial_{i_k}, \qquad \mathbf{B} = \frac{1}{(n-k)!} B_{i_1 \cdots i_{n-k}} dx^{i_1} \wedge \cdots \wedge dx^{i_{n-k}}$$

then the components of **B** will take the form:

$$B_{i_1\cdots i_{n-k}} = \frac{1}{k!}\varepsilon_{i_1\cdots i_{n-k}\,i_{n-k+1}\cdots i_n} A^{i_{n-k+1}\cdots i_n}. \tag{1.25}$$

In particular, if $n = 3$ and $k = 1$ then the 2-form #**X** that is dual to a vector field $\mathbf{X} = X^i \partial_i$ will have the local components:

$$(\#\mathbf{X})_{ij} = \varepsilon_{ijk} X^k. \tag{1.26}$$

If one wishes to define the inverse isomorphism $\#^{-1}: \Lambda^k \to \Lambda_{n-k}$ then it is simplest to first define the reciprocal $n$-vector field **V** to $V$, which must satisfy $V(\mathbf{V}) = 1$. The local form that it takes in a natural frame field is:

$$\mathbf{V} = \partial_1 \wedge \cdots \wedge \partial_n = \frac{1}{n!}\varepsilon^{i_1\cdots i_n}\,\partial_{i_1} \wedge \cdots \wedge \partial_{i_n}. \tag{1.27}$$

The inverse isomorphism $\#^{-1}$ then takes the $k$-form $\alpha$ to the $n-k$-vector field:

$$\#^{-1}\alpha = i_\alpha \mathbf{V}, \tag{1.28}$$

so if $\beta \in \Lambda^{n-k}$ is an $n-k$-form then the evaluation of $\beta$ on $\#^{-1}\alpha$ will equal:

$$\beta(\#^{-1}\alpha) = (\beta \wedge \alpha)(\mathbf{V}). \tag{1.29}$$

This time, the local components of the vector field **X** that is dual to the 2-form $\alpha$ on a three-dimensional manifold will be:

$$X^i = \tfrac{1}{2}\varepsilon^{ijk}\,\alpha_{jk}. \tag{1.30}$$

One can use the Poincaré isomorphism and its inverse to define an adjoint to the exterior derivative operator $d$ by way of:

$$\mathrm{div} = \#^{-1} \cdot d \cdot \#. \tag{1.31}$$

Hence, $\mathrm{div}: \Lambda_k \to \Lambda_{k-1}$.

The notation "div" is suggestive of the divergence operator, and in fact if **X** is a vector field that is locally expressed by $X^i\partial_i$ then tedious, but straightforward, calculations will show that:

$$\mathrm{div}\,\mathbf{X} = \partial_i X^i. \tag{1.32}$$

Furthermore, one has:

$$\mathrm{div} \cdot \mathrm{div} = \#^{-1} \cdot d \cdot \#\#^{-1} \cdot d \cdot \# = \#^{-1} \cdot d \cdot d \cdot \# = 0. \tag{1.33}$$

Hence, one can treat div as a boundary operator, so one can call a $k$-vector field **A** *closed* iff $\mathrm{div}\,\mathbf{A} = 0$ and *exact* iff there is a $k+1$-vector field **B** such that $\mathbf{A} = \mathrm{div}\,\mathbf{B}$. As above, **B** will not be unique, but two such candidates **B**, **B**′ will differ by a closed $k-1$-vector field $\mathbf{z}_{k-1}$:

$$\mathbf{B}' - \mathbf{B} = \mathbf{z}_{k-1} \,. \tag{1.34}$$

One can define the vector spaces $Z_k^{dR}(M)$, $B_k^{dR}(M)$, $H_k^{dR}(M) = Z_k^{dR}(M)/B_k^{dR}(M)$ in complete analogy to the corresponding spaces in de Rham cohomology. In particular the elements of the *de Rham homology vector space in dimension k* $H_k^{dR}(M)$ will be equivalence classes [$\mathbf{z}_k$] of closed *k*-vector fields that differ by exact ones:

$$\mathbf{z}'_k - \mathbf{z}_k = \mathrm{div}\, \mathbf{c}_{k+1} \,. \tag{1.35}$$

Since the definition of div also means that # commutes with div and *d*:

$$\#\, \mathrm{div} = d\, \#, \tag{1.36}$$

the isomorphisms # "descend to homology," and one has corresponding isomorphisms #: $H_k^{dR}(M) \cong H_{dR}^{n-k}(M)$ that are closer to what usually get called the *Poincaré isomorphisms*.

*e. Differentiable homotopies of chains.* If *M* and *N* are differentiable manifolds then a differentiable homotopy between two maps *f*, *g*: $M \to N$ is, quite simply, a differentiable ([1]) map $H : M \times I \to N$, $(x, s) \mapsto H(x, s)$ such that $H(x, 0) = f(x)$ and $H(x, 1) = g(x)$. Each *x* in *M* will then define a differentiable curve in *N* and, by differentiation, a tangent velocity vector field along the curve. However, one also sees that the image of all these curves under *H* will not necessarily foliate the image of *H* into a congruence of curves, although that will be the case if *f* and *g* are diffeomorphisms onto and their images do not intersect. We illustrate some ways that a differentiable homotopy can affect an elementary initial curve in space in Fig. 6.

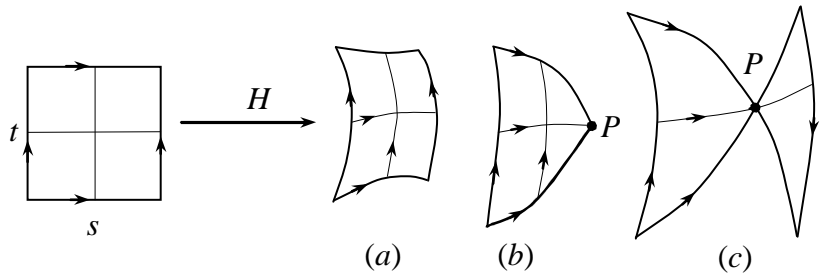

Figure 6.

In order to define a differentiable homotopy of a chain $c_k = \sum_i a^i \sigma_k(i)$ into a chain $c'_k = \sum_i a^i \sigma'_k(i)$, one first defines differentiable homotopies of the basic simplexes $\sigma_k(i)$.

---

([1]) More precisely, in order to define differentiability on the boundary points of $M \times I$, one must extend *H* to a differentiable map from $M \times (-\varepsilon, 1 + \varepsilon)$ and then restrict the extend map to the original points. Since differentiation is local, the choice of extension will be irrelevant to the final result.

Since a differentiable homotopy of a singular $k$-simplex is a singular $(k+1)$-simplex $\sigma_{k+1}$, one then forms the formal sum of the $(k+1)$-simplexes while using the same coefficients $a^i$.

In particular, one defines a differentiable map $H_i: I^{k+1} \to M$, $(s^a, \tau) \mapsto H_i(s^a, \tau) = \sigma_{k+1}(i; s^a, \tau)$ for each $i$ such that $H_i(s^a, 0) = \sigma_k(i)$ and $H_i(s^a, 1) = \sigma'_k(i)$; i.e., each $H_i$ is a $k+1$-simplex $\sigma_{k+1}(i)$ that has $\sigma_k(i)$ and $\sigma'_k(i)$ as components of its boundary, in addition to some "lateral" faces that connect the points of $\sigma_k(i)$ to the corresponding points of $\sigma'_k(i)$.

One then defines the $k+1$-chain $c_{k+1}(0, 1) = \sum_i a^i \sigma_{k+1}(i)$, which has the property that $\partial c_{k+1}(0, 1) = c'_k - c_k + \lambda_k$, in which the $k$-chain $\lambda_k$ represents all of the remaining ("lateral") faces of the boundary that do not cancel out in the sum, and therefore connect points of $\partial c_k$ to corresponding points of $\partial c'_k$. That also means that if the initial $k$-chain $c_k$ is a $k$-cycle $z_k$ then there will be no lateral points and:

$$z'_k - z_k = \partial c_{k+1}(0, 1);$$

i.e., the two $k$-cycles will be homologous. In fact, one can generalize this from the differentiable case to the continuous case, which we then phrase as the:

**Theorem:** *If two singular $k$-cycles are homotopic then they will be homologous.*

This $(k+1)$-chain $c_{k+1}(0, 1)$ is then a differentiable homotopy of $c_k$ into $c'_k$.

From the fact that the exterior derivative operator $d$ commutes with pullbacks of differentiable maps, which implies that $[f^*\alpha] = f^*[\alpha]$, we have a corresponding statement for de Rham cohomology:

**Theorem:** *If $\alpha$ is a closed $k$-form and $f, g: M \to M$ are differentiably homotopic then:* $f^*[\alpha] = g^*[\alpha]$.

(Actually, this property is true for homotopic maps in a more general context than de Rham cohomology.)

If one differentiates $c_{k+1}(0, 1)$ with respect to $\tau$ [i.e., one differentiates each $\sigma_{k+1}(i; s^a, \tau)$] then the resulting velocity vector field will foliate $c_{k+1}(0, 1)$ into a congruence of integral curves, except for possible singular points at which the velocity will be zero. In Fig. 6, (*b*), (*c*), the point *P* would have to be such a point. The restriction of that vector field to the initial $k$-chain $c_k$ can also be regarded as a "variation" of that $k$-chain (cf., e.g., Dedecker [**29**] for this approach to the calculus of variations.)

*f. Chain homotopies.* There is a structure that can be defined in de Rham cohomology that represents the effect of a differentiable homotopy of maps and gets referred to as a "chain homotopy" ([1]), although we shall specialize the definition somewhat. In general, if $f, g : M \to M$ are differentiable maps that are also differentiably homotopic to each other then a *chain homotopy* between them is a linear map $\lambda: \Lambda^k(M) \to \Lambda^{k-1}(M)$ for each $k$ such that:

$$d\lambda + \lambda d = f^* - g^*, \tag{1.37}$$

in which $f^*$, $g^*$ are the pull-back maps in $\Lambda^*(M)$; i.e.:

$$d\lambda\alpha + \lambda d\alpha = f^*\alpha - g^*\alpha \qquad \text{for any } \alpha \in \Lambda^*(M). \tag{1.38}$$

If $\mathbf{u}$ is a vector field on $M$ then $i_\mathbf{u}$ will define such a chain homotopy between two $k$-forms $f^*\alpha$ and $g^*\alpha$ by way of the Lie derivative $L_\mathbf{u}\alpha$ of $\alpha$ with respect to $\mathbf{u}$. From Cartan's formula:

$$L_\mathbf{u} = d i_\mathbf{u} + i_\mathbf{u} d. \tag{1.39}$$

Hence:

$$L_\mathbf{u}\alpha = f^*\alpha - g^*\alpha. \tag{1.40}$$

One also has:

$$d L_\mathbf{u} = L_\mathbf{u} d. \tag{1.41}$$

Moreover, if $\alpha$ is closed then:

$$L_\mathbf{u}\alpha = d i_\mathbf{u} \alpha, \tag{1.42}$$

which is exact. An obvious corollary to this is that if $\alpha$ is exact then so is $L_\mathbf{u}\alpha$.

We can then state:

**Theorem:** *If $\alpha$ is closed (and thus defines a de Rham cohomology class $[\alpha]$) then $[L_\mathbf{u}\alpha] = 0$.*

As a partial converse to this, if $L_\mathbf{u}\alpha = d\beta$ for some $\beta$ then:

$$i_\mathbf{u} d\alpha = d(\beta - i_\mathbf{u} \alpha), \tag{1.43}$$

which is exact, although that does not imply that $\alpha$ is exact.

**2. Vector homology.** The way that homology is often introduced (if only implicitly) into the study of motion is by considering the motion of a $k$-chain along the flow of a vector field $\mathbf{v}$. Perhaps the simplest such situation that one encounters in fluid motion is the convection of a fluid cell (i.e., a $k$-simplex $\sigma_k$) along the flow of the flow velocity vector field $\mathbf{v}$, which is depicted in Fig. 7.

---

([1]) Under the circumstances, one should probably say *cochain* homotopy, but the present terminology is so widespread (cf., e.g., [**9**, **30**]), it would only confuse people to change it now.

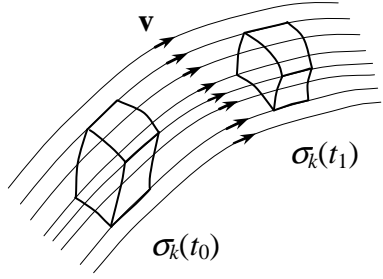

Figure 7. Convection of a basic fluid cell (i.e., simplex).

More generally, one can regard the $k+1$-dimensional space that is swept out by a $k$-chain $c_k$ at different times $t_0$, $t_1$ in its motion as a $k+1$-chain $c_{k+1}(t_0, t_1)$ whose boundary consists $\partial c_{k+1}(t_0, t_1) = c_k(t_0) - c_k(t_1)$. Hence, $c_k(t_0)$ and $c_k(t_1)$ must be cobounding. When the $k$-chains are $k$-cycles, $c_{k+1}(t_0, t_1)$ then becomes a homology between the $k$-cycles. However, it is more specific than a general homology that one encounters in singular homology, since there is also a one-dimensional foliation of it by the integral curves of some vector field **u**, whose support will typically be larger than the image of $c_{k+1}$.

Hence, we define a *(singular) vector homology* ([1]) between a two cobounding $k$-chains $c_k$ and $c'_k$ to be a differentiable $k+1$-chain $c_{k+1}$ and a vector field **u** on such that:

1. $\partial c_{k+1} = c'_k - c_k$.

2. The vector field **u** that is defined on $c_{k+1}$ and intersects $c_k$ and $c'_k$ transversally, such that **u** is interior normal to $c_k$ and exterior normal to $c'_k$.

This definition differs from Reinhardt's definition of vector cobordism by the fact that he also requires **u** to have no zeroes on $c_{k+1}$. The reason that we are including the singular case of vector fields with zeroes is because it allows for the formation of vortex pairs from a single point, which will have to be a zero of the flow velocity vector field.

Note that there is nothing to say that the two boundary components $c_k$ and $c'_k$ are connected as topological spaces. In fact, they might well be represented by $k$-simplexes whose boundaries are not identified in any way by $\partial$.

Since we are assuming differentiability of $c_{k+1}$, the integral curves of **u** will foliate $c_{k+1}$ with a congruence when **u** has no zeroes. We illustrate a typical non-singular situation schematically in Fig. 8:

---

([1]) This is a more specialized form of what Reinhardt calls "vector cobordism [**31**]" and what physicists call "Lorentz cobordism" when $k = 3$. However, the scope of cobordism, as a generalized homology, is more appropriate to purely mathematical problems that involve the classification of differentiable manifolds, but at the expense of providing tools for modeling the more mundane phenomena that nature presents in a laboratory. Hence, we shall introduce a specialization of the concept to the more easily-computed world of singular homology.

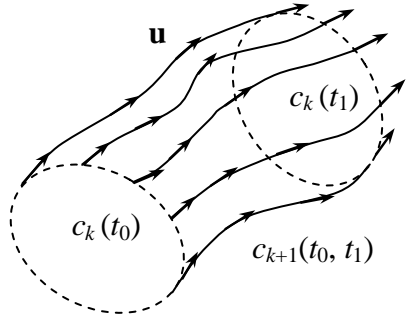

Figure 8. Typical vector homology between *k*-chains.

However, this kind of homology also includes the possibility of "topology-changing processes," such as when the chain (which must be a cycle) collapses to a point *P*, which must then be a singular point of **u**, or when it splits into two disjoint copies of itself. We then illustrate these possibilities in Fig. 9.

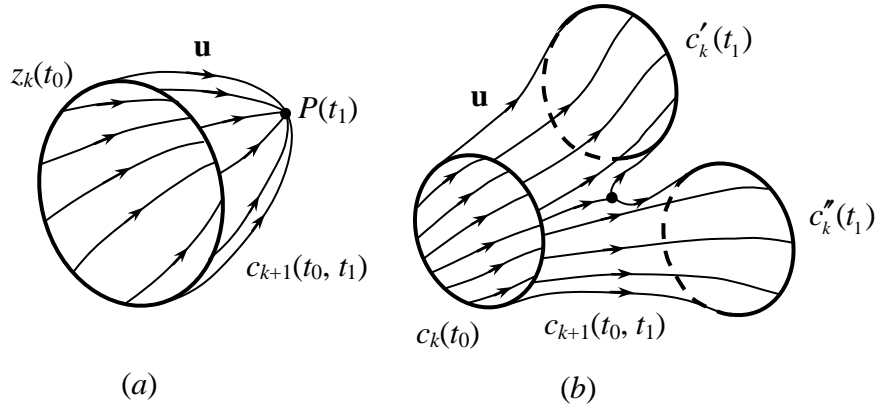

Figure 9. Topology-changing processes as vector homologies.

Note that in case (*a*), one can say $z_k(t_0)$ is also homologous to zero; i.e., it is a boundary, which is why it must be a cycle, as well. Furthermore, the point $P(t_1)$ must be a zero of the vector field **u**, and one sees that in the illustration, it will take the form of an attractor. In the case (*b*), **u** will typically have a zero at a point in the "crotch," but it will be a saddle point; i.e., it will be attracting in one planar section and repelling in another.

Since a vector homology involves the use of a vector field in addition to an interpolating chain, one sees that two homologies – i.e., two *k*+1 chains – that are cobounding might still involve vector fields that foliate them in "inequivalent" ways. In order to define that stronger notion of equivalence for vector homologies, we take the position that the congruences of curves define the most useful kind of equivalence. A natural equivalence for congruences is "concordance ([1])," which amounts to saying that each of the curves of one congruence is homotopic to some distinct curve of the other, although we shall phrase this in terms of the vector field **u**.

---

([1]) See, for instance, the way that this is introduced in Kuschorke [**32**].

Let $\hat{c}_{k+1}$ be the set of points of $c_{k+1}$ at which **u** is non-zero. One can then normalize **u** to a unit vector field $\hat{\mathbf{u}}$ at those points, if one introduces some auxiliary metric. (Actually, it is topologically unnecessary to introduce one, but it makes things simpler.) A unit vector field on $\hat{c}_{k+1}$ can be regarded as a section of a tangent sphere bundle $S(\hat{c}_{k+1})$ whose fibers are the unit spheres in each fiber of $T(\hat{c}_{k+1})$. One can then talk about homotopies of sections – i.e., unit vector fields on $\hat{c}_{k+1}$ – and that is how we will define concordance of congruences; i.e., by the homotopy of their unit vector fields at the non-singular points.

Hence, our definition of equivalence for vector homologies $(c_{k+1}, \mathbf{u})$ and $(c'_{k+1}, \mathbf{u}')$ between given $k$-chains is that $c_{k+1}$ and $c'_{k+1}$ must be cobounding, while **u** and **u**′ must have the same number of zeroes and their unit vector fields must be homotopic where they are defined. An example of inequivalent vector homologies between two $k$-chains is given by comparing Fig. 8 (with $k = 1$) with Fig. 10. One sees that in both cases, $c_2(t_0, t_1)$ is a 2-cylinder, but the vector field **u** has no zeroes in Fig. 8, while it has two in Fig. 10.

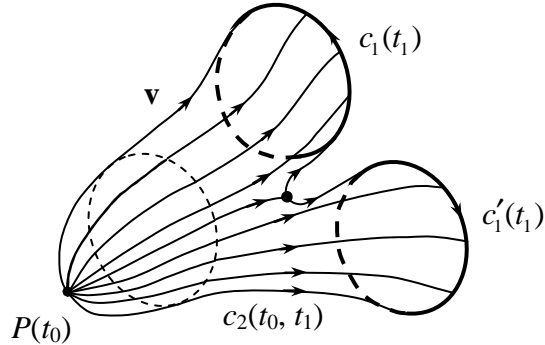

Figure 10. The creation of a vortex-pair from a point.

It is perhaps in the context of fluid mechanics that one can get the best intuition for how topology-changing processes occur in nature, since that branch of theoretical mechanics offers some of the most familiar examples. For instance, one can have the nucleation or cavitation of bubbles in fluids, and the formation of smoke rings, which are described by the illustration in Fig. 9 (*a*) for various dimensions $k$, and the formation of vortex-pairs in turbulent fluids, which is more akin to the illustration in Fig. 10. One can also get the 2-chain $c_2(t_0, t_1)$ by combining the homologies in Figs. (*a*) and (*b*) and identifying common points on their boundaries.

A useful example of a vector homology is given by any differentiable homotopy of a $k$-chain $c_k(0) = \sum_i a^i \sigma_k(i)$ to a $k$-chain $c_k(1)$.

**3. Integral invariants.** One can then see that vector homologies are perhaps the best way of introducing the concept of "integral invariants," at least in the context of the topology of motion. Basically, an *integral invariant* [**33, 34**] is an integral that keeps the same numerical value as one follows the flow of some motion. Since we know that integrals are best defined when one has a $k$-form and a $k$-chain, we then introduce a differential *k*-form $\alpha$ into the picture of vector homology that we can integrate over the boundary components and differentiate on the *k*+1-chain.

In the absence of the vector field **u**, in order to make the integral of $\alpha$ the same over both boundary components in Fig. 8, it is sufficient for $\alpha$ to be closed:

$$0 = \int_{c_k(t_1)} \alpha - \int_{c_k(t_0)} \alpha = \int_{c_k(t_1) - c_k(t_0)} \alpha = \int_{\partial c_k(t_0, t_1)} \alpha = \int_{c_k(t_0, t_1)} d\alpha.$$

If that is true for all $c_k$ then it will also be necessary.

Hence, the integral in question would define a singular *k*-cocycle. In that light, an integral invariant would be essentially a homology-invariant. The generalization from Fig. 8 is then based in the idea that $\partial c_{k+1}(t_0, t_1)$ can be composed of more than two components, or even less than two. In the latter case, it is necessary and sufficient that $\alpha$ must be exact, since its integral over any *k*-cycle must vanish.

If we introduce the vector field **u** and require that the boundary components must be transverse to the integral curves then we must restrict the *k*-form $\alpha$ by more than just the vanishing of its exterior derivative. In particular, we also demand that the integral of $\alpha$ must be the same over *all* intermediate *k*-chains $c_k(t)$ that result from allowing the points of $c_k(t_0)$ to flow along the integral curves of **u**, and not just the "initial" and "final" ones.

Hence, let us revert to Fig. 8 and replace $c_k(t_0)$ with $c_k(t)$, and $c_k(t_1)$ with $c_k(t + \Delta t)$, which is then the convected version of $c_k(t)$ along the flow of **u** after a time interval of $\Delta t$. We also assume that since we will be looking at the limit as $\Delta t$ goes to zero, we can consider only the immediate neighborhood of $c_k(t)$, in which there will be a one-parameter group of diffeomorphisms onto $\Phi_{\Delta t} : c_k(t) \to M$, so $c_k(t + \Delta t)$ will also be diffeomorphic to $c_k(t)$; i.e., topology-changing processes will not be an issue.

We define the integral of $\alpha$ over $c_k(t)$ for any *t* to be the function of time:

$$C(t) = \int_{c_k(t)} \alpha. \tag{3.1}$$

If this integral is constant along the flow of **u** for any *k*-chain $c_k(t)$ then we will call $C(t)$ an *absolute integral invariant*.

Since $c_k(t)$ is assumed to be differentiable, $C(t)$ will also be differentiable, and:

$$\frac{dC}{dt} = \lim_{\Delta t \to 0} \frac{1}{\Delta t} \left[ \int_{c_k(t+\Delta t)} \alpha - \int_{c_k(t)} \alpha \right] = \int_{c_k(t)} \lim_{\Delta t \to 0} \frac{1}{\Delta t} (\Phi_{\Delta t}^* \alpha - \alpha) = \int_{c_k(t)} L_{\mathbf{u}} \alpha.$$

That is ([1]):

$$\frac{dC}{dt} = \int_{c_k(t)} L_{\mathbf{u}} \alpha. \tag{3.2}$$

Hence, if this is to be true for all $c_k(t)$ then:

---

([1]) The details of this calculation are given in Chap. VIII of Godbillon [**34**].

**Theorem:** *C is an absolute integral invariant for the flow of* **u** *iff:*

$$L_{\mathbf{u}}\alpha = 0. \tag{3.3}$$

One can substitute a weaker condition than this when one restricts the *k*-chain $c_k(t)$ to be a *k*-cycle $z_k(t)$. Although the vanishing of $L_{\mathbf{u}}\alpha$ would still be sufficient for the vanishing of $dC / dt$, it would not be necessary, since $L_{\mathbf{u}}\alpha = i_{\mathbf{u}}d\alpha + di_{\mathbf{u}}\alpha$, and the integral of the second term will vanish when one integrates it over a *k*-cycle. Hence, *C* will be constant along the flow of **u** for any such $z_k(t)$ iff:

$$i_{\mathbf{u}}d\alpha = d\beta \tag{3.4}$$

for some *k*−1-form $\beta$.

Now:

$$dL_{\mathbf{u}}\alpha = L_{\mathbf{u}}d\alpha, \tag{3.5}$$

so one sees that if $L_{\mathbf{u}}\alpha$ is closed then $d\alpha$ will define an absolute integral invariant ([1]). We shall then call *C*(*t*) a *relative integral invariant* in such a case. Hence:

**Theorem:** $C = \int_{c_k} \alpha$ *is a relative integral invariant of the flow of* **u** *iff* $[L_{\mathbf{u}}\alpha] = 0$*; i.e.:*

$$L_{\mathbf{u}}\alpha = d\beta \quad \text{(for some } k{-}1\text{-form } \beta\text{).} \tag{3.6}$$

Clearly, demanding that $L_{\mathbf{u}}\alpha$ must vanish is a stronger condition than requiring that it must merely be exact.

From (3.2) and (3.5), we have the obvious:

**Corollary:** *Any de Rham cohomology class* $[\alpha]$ *in dimension k defines a relative integral invariant for the flow of any vectorfield.*

(Namely, $d\alpha = 0$ implies that $L_{\mathbf{u}}\alpha = di_{\mathbf{u}}\alpha$ is exact for any **u**.) This corollary is true because cohomology classes are invariant under the more general homologies that do not depend upon a choice of vector field.

Furthermore, since any *n*-form on an *n*-dimensional manifold is closed, we have the:

**Corollary:** *Any n-form on an n-dimensional manifold defines a relative integral invariant for the flow of any vector field.*

However, the question of whether these relative integral invariants are also absolute ones is not as guaranteed:

---

[1] We caution the reader that since we are defining integral invariants by starting with the integrals themselves, we are imposing stronger conditions on $\alpha$ than, say, Godbillon [34], who was following Cartan [33]. For him, an absolute integral invariant is a *k*-form $\alpha$ such that $i_{\mathbf{u}}\alpha$ and $i_{\mathbf{u}}d\alpha$ both vanish, while a relative integral invariant is an $\alpha$ such that $d\alpha$ is an absolute integral invariant; hence, it is necessary and sufficient that $i_{\mathbf{u}}d\alpha$ must vanish.

**Theorem:** *A relative integral invariant $\alpha$ is also an absolute integral invariant iff $i_\mathbf{u}\alpha$ is closed.*

Hence, $i_\mathbf{u}\alpha$ will define a de Rham cohomology class $[i_\mathbf{u}\alpha]$ in dimension $k - 1$ in that event.

One should note that equation (3.4) can also be interpreted by saying that there is a chain homotopy between the exact form $d\beta$ and 0.

We shall use this basic construction of vector homology later in the context of flow tubes and vortex tubes, which are examples of vector homologies when the vector field is the flow velocity vector field or the vorticity vector field, respectively. We shall then see that the main theorems of vortex motion take the form of asserting that some of the fundamental integrals of that theory define certain integral invariants of the motion.

Part II: The theory of vortices

**4. Basic fluid kinematics.** – The fluid state of matter is characterized by the property (among others) that the deformation (i.e., strain) of a volume of the material does not produce any resulting stress, except for an overall pressure that couples to the dilatation. However, the *rate* of strain can induce stresses in the form of viscosity. Hence, the story of fluid kinematics usually starts with the flow velocity, rather than the actual deformation of the fluid. (Basically, one is using the "Euler" picture of continuum motion, rather than the "Lagrange" picture.)

*a. The basic flow region.* Let us assume that our fluid occupies a region *M* of space, which we then assume to be an *n*-dimensional differentiable manifold ($n = 1, 2,$ or $3$, typically) that will often have a boundary $\partial M$, such as for flows in channels; the case without a boundary might perhaps relate to clouds and nebulae. In fact, the boundary might be sufficiently piecewise-linear that only the interior points of *M* can be regarded as a manifold. In order to make it easier to deal with such cases, we will assume that *M* is a subset of $\mathbb{R}^n$, so the points where *M* might not admit a legitimate tangent space will simply be regarded as points of $\mathbb{R}^n$.

The focus of this article is on topologically non-trivial flow regions, as opposed to trivial cases such as discs and balls. Examples of topologically non-trivial manifolds *M* that fluid mechanics often presents are annular channels between circular walls, fluid shells between spherical walls, and toral regions of flow.

Time will take the form of a point on either the real line $\mathbb{R}$ or a finite interval $(t_0, t_1)$ along $\mathbb{R}$. In the case of non-relativistic motion, time will also double as a universal curve parameter for all motion in *M* and a dimension of space-time, so our non-relativistic space-time will take the form of either $\mathbb{R} \times M$ or $(t_0, t_1) \times M$, depending upon the time interval that is specified. Hence, in either case, the homology of the space-time will still be due to *M*.

Typically, the manifold *M* will also be assumed to be orientable and Riemannian. Hence, as discussed above, *M* will be endowed with a volume element $V \in \Lambda^n(M)$, which is a global, non-zero *n*-form on *M*. *V* will then define the Poincaré isomorphisms $\# : \Lambda_k \to \Lambda^{n-k}$, $\mathbf{B} \mapsto \#\mathbf{B}$, where $\#\mathbf{B} = i_\mathbf{B} V$.

To say that *M* is *Riemannian* is to say that each of its tangent spaces $T_x M$ is endowed with a positive-definite scalar product $g_x(\mathbf{v}, \mathbf{w})$ that will locally look like:

$$g = g_{ij}(x)\, dx^i\, dx^j, \tag{4.1}$$

in which the multiplication of 1-forms is symmetrized tensor multiplication.

A local coframe field $\{\theta^i, i = 1, \ldots, n\}$ on *M* is *orthonormal* iff the components of *g* with respect to that coframe field are $\delta_{ij}$:

$$g = \delta_{ij}\, \theta^i\, \theta^j. \tag{4.2}$$

*b. Flow velocity vector field.* The *flow velocity* vector field on the space-time region that is occupied by a fluid takes the form **v**(*t*, *x*), in general. Thus, we are starting with a *time-varying* vector field. In the event that **v** = **v**(*x*), the flow will be referred to as *steady*, and since *t* plays no role, one will usually deal with *M* directly. More generally, the steady-flow regime of fluid motion will be characterized by assuming that all of the fluid state variables are time-independent, and not just flow velocity; in particular, the mass density and pressure will also be functions of only the points of space.

Locally, the flow velocity vector field will take the form in a natural frame field $\{\partial_i = \partial/\partial x^i, i = 1, \ldots, n\}$:

$$\mathbf{v}(t, x) = v^i(t, x)\, \partial_i, \qquad (4.3)$$

and for steady flow, the components of **v** take the functional form $v^i(x)$.

It will often be useful to regard the component of **v** that goes with $\partial/\partial t$ as 1, which would make:

$$\mathbf{v}(t, x) = \partial_t + \mathbf{v}_s = \partial_t + v^i(t, x)\, \partial_i. \qquad (4.4)$$

For each value of *t*, the flow region *M* will be foliated by a congruence of integral curves of **v**(*t*, *x*) that one calls *path lines* in the general, time-varying case. That is, a point *x*(*t*) along such a curve will satisfy:

$$\frac{dx}{dt} = \mathbf{v}(t, x). \qquad (4.5)$$

Hence, as the flow velocity vector field varies in time, so will the congruence of path lines.

In the steady flow case, the congruence will stay spatially fixed as time elapses, and one will call the integral curves *streamlines*.

A common construction involving **v** is that of a *flow tube*, which basically amounts to a vector homology of the kind that was depicted in Fig. 8. That is, one has a homology $c_2(0, 1)$ between two 1-chains $c_1(0)$ and $c_1(1)$ such that the restriction of **v** to $c_2(0, 1)$ is transverse to its boundary. Hence, $c_2(0, 1)$ will also be foliated by the path lines or streamlines of **v**.

*c. The velocity gradient.* The differential *d***v** of the flow velocity vector field – viz., the *velocity gradient* – contains the next set of essential kinematical information that pertains to fluid flow. If its components relative to the natural frame are:

$$v^i_{,t} = \frac{\partial v^i}{\partial t} \qquad \text{and} \qquad v^i_{,j} = \frac{\partial v^i}{\partial x^j} \qquad (4.6)$$

then

$$d\mathbf{v} = v^i_{,t}\, dt \otimes \partial_i + v^i_{,j}\, dx^j \otimes \partial_i. \qquad (4.7)$$

The matrix $v^i_{,j}$ represents an infinitesimal linear transformation of $\mathbb{R}^n$. One can polarize it with respect to the transpose operator and get:

$$v^i_{,j} = \tfrac{1}{2}(\dot{e}^i_j + \omega^i_j), \tag{4.8}$$

in which we have defined the components of the *rate of strain* tensor:

$$\dot{e}^i_j = v^i_{,j} + v^j_{,i} \tag{4.9}$$

and the components of the *kinematical vorticity* tensor:

$$\omega^i_j = v^i_{,j} - v^j_{,i}. \tag{4.10}$$

The matrix $\dot{e}^i_j$ is then symmetric and can be further reduced to a part $\hat{e}^i_j$ that has zero trace and a trace part:

$$\dot{e}^i_j = \hat{e}^i_j + \dot{e}\,\delta^i_j, \tag{4.11}$$

in which:

$$\hat{e}^i_j = \dot{e}^i_j - \dot{e}\,\delta^i_j, \qquad \dot{e} = \frac{1}{n} v^k_{,k}. \tag{4.12}$$

If one recalls the definition of the (spatial) divergence of a $k$-vector field **B** on $M$ from above then one will have:

$$\text{div } \mathbf{v} = \#^{-1} d_s \# \mathbf{v} = v^k_{,k}. \tag{4.13}$$

Hence, the trace part of $d\mathbf{v}$ is proportional to the divergence of **v**, which one calls the *compressibility* of the flow of **v**, since when div **v** = 0, one calls the flow *incompressible* and *compressible* otherwise. A vector field with vanishing divergence is the infinitesimal generator of a one-parameter family of volume-preserving diffeomorphisms. One can see this easily by calculating the Lie derivative of $V$ along the flow of **v**:

$$\mathbf{L}_\mathbf{v} V = i_\mathbf{v} d_s V + d_s i_\mathbf{v} V = d_s \# \mathbf{v} = \#^{-1}(\text{div } \mathbf{v}), \tag{4.14}$$

so $\mathbf{L}_\mathbf{v} V$ will vanish iff div **v** does.

Notice that since (4.14) includes the fact that:

$$\mathbf{L}_\mathbf{v} V = d_s \# \mathbf{v}, \tag{4.15}$$

one can say that for steady flow, the spatial volume element defines a relative integral invariant for the flow of **v**, even for compressible flow, and it will be an absolute integral invariant iff the flow is incompressible.

The other part of $v^i_{,j}$ – namely, $\omega^i_j$ – is the infinitesimal generator of a one-parameter family of rotations of $\mathbb{R}^n$. In fact, it can be regarded as a position-dependent angular velocity for the local motion of the fluid. It is related to the (kinematical) vorticity 2-form, which shall be the basis for the next section of this study.

*d. Flow covelocity 1-form.* For many purposes, it is more convenient to deal with the *covelocity 1-form:*

$$v = dt + v_i(t, x)\, dx^i \tag{4.16}$$

that one generally associates with the velocity vector field **v**, as it is defined in (4.4), by using the linear isomorphism of spatial tangent spaces and cotangent spaces that is defined by the metric tensor *g*:

$$v_s = i_{\mathbf{v}_s} g \qquad [\text{i.e., } v_s(\mathbf{w}) = g(\mathbf{v}_s, \mathbf{w})], \tag{4.17}$$

so the spatial components will be converted by the usual "lowering of the index":

$$v_i = g_{ij}\, v^j. \tag{4.18}$$

Hence, just as a non-zero velocity **v** spans a line in each tangent space, a non-zero covelocity *v* will define a hyperplane in that tangent space, namely, the annihilating hyperplane of *v*. Since that amounts to all tangent vectors **w** such that $v(\mathbf{w}) = 0$, and that also implies that they all satisfy $g(\mathbf{v}_s, \mathbf{w}_s) = 0$, one can characterize the hyperplane that *v* defines as being orthogonal to the line that **v** defines.

*e. Integrability of covelocity.* The question of the integrability of the differential system on *M* that is defined by all of the annihilating hyperplanes to *v* − which one expresses by saying that they are the (algebraic) solution to the *exterior differential system* $v = 0$ – is much more involved than the integrability of the one-dimensional differential system that is defined by a non-zero **v**. An analytical solution to that same system, which was originally called the *Pfaff problem* [**35-38**], is a submanifold of *M* whose tangent spaces are subspaces of the annihilating hyperplanes of *v*. The dimension of such a submanifold can potentially range from one to $n - 1$ (i.e., codimension one) and is referred to as the *degree of integrability* of the differential system. One-dimensional integral submanifolds (i.e., integral curves) always exist, but the existence of codimension-one integral submanifolds is referred to as *complete integrability*, which does not have to occur.

There are two ways that this can happen, as far as *v* is concerned: Either *v* is an exact form or it admits an integrating factor; i.e.:

$$v = d\psi \qquad \text{or} \qquad v = \lambda\, d\mu. \tag{4.19}$$

In the first case, the integral submanifolds are level hypersurfaces of the function $\psi$, and in the second case, they are level hypersurfaces of $\mu$.

If *v* is exact then the function $\psi$ (which is, of course, defined only up to an arbitrary constant on each connected component of *M*) is commonly called the *stream function* or *velocity potential* of the flow of **v**. In the second case, $1/\lambda$ is referred to as an "integrating factor" for the 1-form *v*, since $(1/\lambda)\, v$ would be exact – i.e., completely integrable.

A necessary condition for the first case is that *v* must be closed ($dv = 0$); whether that is also sufficient will depend upon whether *M* is simply-connected or not, respectively.

We shall elaborate upon this issue later in this article, but for now, we observe that in the second case of *v*, if one takes the exterior derivative of *v* then one will get $d\lambda \wedge d\mu$, and although this will vanish iff $d\lambda$ is proportional to $d\mu$ by way of a non-zero scalar function (which is then a degenerate condition), nonetheless, one will always have:

$$\mathfrak{F} \equiv v \wedge dv = 0. \tag{4.20}$$

One notices that both possibilities in (4.19) satisfy that condition, and in fact, one version of *Frobenius's theorem* is that $v = 0$ is completely integrable iff $\mathfrak{F} = 0$; we shall then call $\mathfrak{F}$ the *Frobenius 3-form* that is defined by *v*.

When *M* has a dimension of two, $\mathfrak{F}$ must always vanish, and every *v* will be completely integrable. When *M* has dimension three, $\mathfrak{F}$ can still be non-zero, since we have assumed orientability. In such a case, there is one more possible form that *v* can take, in addition to the ones in (4.19), namely:

$$v = d\psi + \lambda\, d\mu. \tag{4.21}$$

In this case, one will generically have:

$$\mathfrak{F} = d\psi \wedge d\lambda \wedge d\mu, \tag{4.22}$$

which will vanish iff the 3-coframe $\{d\psi, d\lambda, d\mu\}$ degenerates to something linearly-dependent.

However, if *M* is three-dimensional then $dv \wedge dv$ must vanish identically, since it is a 4-form. Of course, that 4-form will still be potentially non-trivial when one goes to relativistic fluid mechanics, or unsteady, non-relativistic flow, which we shall discuss below.

In the three-dimensional case with steady flow, when $\mathfrak{F} = 0$, one also says that the flow of **v** is *surface-orthogonal*, since the tangent subspaces that are defined by $v = 0$ are planes that are orthogonal to the tangent lines that are spanned by **v**. If $\mathfrak{F} \neq 0$ then the only orthogonal integral submanifolds will be curves, not surfaces.

There is a basic sequence of canonical forms that dictates the degree of integrability of $v = 0$ that is defined by:

$$\mathcal{I}_0 = v, \quad \mathcal{I}_1 = dv, \quad \mathcal{I}_2 = v \wedge dv, \quad \mathcal{I}_3 = dv \wedge dv, \quad \mathcal{I}_4 = v \wedge dv \wedge dv, \ldots$$

In general, one has the recursion:

$$\left. \begin{array}{l} \mathcal{I}_{2k-1} = d\mathcal{I}_{2k}, \quad k = 0, 1, \ldots \\ \mathcal{I}_{2k} = v \wedge \mathcal{I}_{2k-1}, \quad k = 1, 2, \ldots \end{array} \right\} \tag{4.23}$$

The first of these differential forms to vanish globally on *M* will dictate the degree of integrability of $v = 0$. We have already seen that if $\mathcal{I}_1$ or $\mathcal{I}_2$ is the first to vanish (i.e., $k = 1$) then the integral submanifolds will have codimension one, and thus the degree of integrability will be $n - 1$. If $\mathcal{I}_3$ is the first to vanish (i.e., $k = 2$) then the integral submanifolds will have codimension two, and thus the degree of differentiability will be $n - 2$. The other possibilities relate to manifolds of dimension higher than three, and will thus be revisited below in the context of unsteady flow, as well as in the sequel to this article that will be concerned with relativistic considerations. In general, though, if the first form that vanishes globally corresponds to some value of $k$ then the degree of integrability will be $n - k$.

One notices that the forms $\mathcal{I}_{2k-1}$ are all exact, and thus define trivial de Rham cohomology classes in their dimensions; i.e., $[\mathcal{I}_{2k-1}] = 0$. By contrast, whether the even forms are closed, and thus define de Rham cohomology classes $[\mathcal{I}_{2k}]$, will depend upon whether $\mathcal{I}_{2k-1} = 0$; whether the Pfaff problem $v = 0$ has degree of integrability $n - k$. In fact:

**Theorem:** *If the degree of integrability of the Pfaff problem is $n - k$ then $\mathcal{I}_{2k}$ will be closed; i.e., it will define a de Rham cohomology class in dimension $n - k$.*

Whether $\mathcal{I}_{2k}$ is also exact will then depend upon the dimension of the de Rham cohomology vector space of $M$ in dimension $2k + 1$. In particular, if $M$ is three-dimensional and orientable, since $H^3(M)$ must contain a fundamental, non-zero 3-cocycle $[V]$, that dimension will be greater than zero, and $\mathcal{I}_2 = v \wedge dv$ will be closed, but not necessarily exact.

*e. Convective acceleration 1-form.* We next examine the nature of what one could refer to as the *convective acceleration*, namely:

$$a = \mathbf{L_v} v = i_{\mathbf{v}} dv + d i_{\mathbf{v}} v. \tag{4.24}$$

It would then describe the time rate of change of the covelocity 1-form $v$ along the flow of $\mathbf{v}$.

If we represent $\mathbf{v}$ as $\partial_t + \mathbf{v}_s$ and $v$ as $dt + v_s$ then direct calculation will give:

$$a = \frac{dv_s}{dt} + \tfrac{1}{2} dv^2, \tag{4.25}$$

in which we have introduced the more traditional *convected derivative* of the covelocity:

$$\frac{dv_s}{dt} = \frac{dv_i}{dt} dx^i, \qquad \frac{dv_i}{dt} \equiv \partial_t v_i + v^j \partial_j v_i. \tag{4.26}$$

In local components, we also have:

$$dv^2 = (\partial_t v^2)\, dt + d_s v^2. \tag{4.27}$$

**5. Kinematical vorticity 2-form.** – We shall now devote more specific attention to the 2-form $dv$ that was introduced above.

*a. Basic definition.* The 2-form:

$$\Omega = dv \tag{5.1}$$

is referred to as the *kinematical vorticity* 2-form of the flow of **v**; had we started with the *momentum density* 1-form $p = \rho v$, instead of $v$, then the resulting 2-form:

$$\Omega_d = dp = d\rho \wedge v + \rho\,\Omega \tag{5.2}$$

would be called the *dynamical vorticity*; since we shall be primarily concerned with the kinematical case, we shall not specify it with a subscript. Clearly, $\Omega_d$ is proportional to $\Omega$ iff $d\rho \wedge v$ is proportional to $\Omega$, such as when $\rho$ is spatially constant. However, in what follows, we shall use the term "vorticity" when we are referring to the kinematical kind as the default case.

If $\Omega = 0$ then one will refer to the flow of **v** as *irrotational*, and otherwise, *vorticial*. Hence, irrotational flows will define a 1-dimensional de Rham cohomology class $[v]$. Whether it vanishes will depend upon whether $M$ is simply-connected, and we shall return to that question shortly.

One sees that when $v$ does not correspond to steady flow, the 2-form $dv$ will have to be defined on the space-time manifold $\mathbb{R} \times M$, since one will have to differentiate with respect to time. In time+space form, it will look like.

$$\Omega = dt \wedge \partial_t v_s + \omega, \tag{5.3}$$

in which we have defined the partial derivative of $v_s$ with respect to time and the spatial vorticity 2-form as:

$$\partial_i v_s = (\partial_i v_s)\,dx^i, \qquad \omega = \tfrac{1}{2}(\partial_i v_j - \partial_j v_i)\,dx^i \wedge dx^j. \tag{5.4}$$

It is the latter spatial 2-form whose components are what usually get referred to as the vorticity in conventional treatments of fluid mechanics. For steady flow, $\partial_t v_s = 0$, and one can simply define $\omega = dv$ to be the vorticity on $M$.

We find that the Frobenius 3-form is:

$$\mathfrak{F} = v \wedge \Omega = dt \wedge (\omega + \partial_t v_s \wedge v_s) + v_s \wedge \omega, \tag{5.5}$$

which will vanish iff:

$$\omega = -\partial_t v_s \wedge v_s, \qquad v_s \wedge \omega = 0. \tag{5.6}$$

We can also calculate:

$$\Omega \wedge \Omega = 2\,dt \wedge \partial_t v_s \wedge \omega, \tag{5.7}$$

which will vanish iff:

$$\partial_t v_s \wedge \omega = 0. \tag{5.8}$$

If one refers back to the expression (4.10) for $\omega_j^i$ then one will see that if one lowers the upper index using $g$ then the spatial components of $\Omega_k$ will be equal to $-\omega_{ij}$.

In the case of steady, 2-dimensional flow, for which $v$ will take the form $v_1 \, dx^1 + v_2 \, dx^2$, the kinematical vorticity 2-form will be simply:

$$\Omega = (\partial_1 v_2 - \partial_2 v_1) \, dx^1 \wedge dx^2. \tag{5.9}$$

As long as the flow channel is orientable, this does not have to vanish.

From the previous discussion of the Pfaff problem that is defined by $v$, we see that for vorticial flows in three-dimensional spaces, the degree of integrability will come down to whether $v$ takes the form $\lambda \, d\mu$ or $d\psi + \lambda \, d\mu$. In the latter case, the functions $\lambda$ and $\mu$ are referred to as the *Clebsch parameters* [**19, 39**].

One can also express $dv/dt$ in terms of the kinematical vorticity 2-form $\omega$ by going back to (4.24) and taking (5.3) into account, which will then make:

$$\frac{dv}{dt} = (\partial_t v_i - \omega_{ij} v^j + \tfrac{1}{2}\partial_i v^2) \, dx^i. \tag{5.10}$$

*b. Vorticity vector field.* One can use the inverse Poincaré isomorphism $\#^{-1}: \Lambda^2 \to \Lambda_0$ when $M$ is two-dimensional and $\#^{-1}: \Lambda^2 \to \Lambda_1$ when $M$ is three-dimensional to define dual objects $\boldsymbol{\omega} = \#^{-1}\omega$ to the spatial vorticity 2-form $\omega$ in each case.

In the two-dimensional case, the dual object will be a scalar field whose local form will be, from (5.9):

$$\omega = \frac{\partial v_2}{\partial x^1} - \frac{\partial v_1}{\partial x^2}. \tag{5.11}$$

In the three-dimensional case, the dual object will be a vector field that one calls the *vorticity vector field*, and whose local (spatial) components will take the form:

$$\omega^i = \tfrac{1}{2}\varepsilon^{ijk}(\partial_j v_k - \partial_k v_j), \tag{5.12}$$

which, in fact, agree with the components of $\tfrac{1}{2}\text{curl } \mathbf{v}$.

An immediate consequence of the definition of $\boldsymbol{\omega}$ (in three dimensions) is:

**Theorem:** *The flow of the vorticity vector field $\boldsymbol{\omega}$ is incompressible*.

The proof of this follows by direct calculation:

$$\text{div } \boldsymbol{\omega} = \#^{-1} d_s \# \#^{-1} \omega = \#^{-1} d_s \, d_s v_s = 0.$$

Hence, [$\boldsymbol{\omega}$] defines a de Rham homology class in dimension one that will, however, vanish, since:

**Theorem:** *There exists a bivector field* **B** *such that* $\omega = \text{div } \mathbf{B}$.

The proof follows from the fact that:

$$\omega = \#^{-1} d_s v = \#^{-1} d_s \, \# \, \#^{-1} v = \text{div}(\#^{-1} v).$$

Hence, it is sufficient to set $\mathbf{B} = \#^{-1} v_s$.

The last two theorems are actually dual to the fact that $\omega = d_s v_s$ is exact, and therefore closed, and the fact that the isomorphisms # "descend to homology"; i.e., they commute with $d_s$ and div.

c. *Vortex lines and tubes.* Like any vector field on $M$, $\omega$ defines a system of first-order ordinary differential equations:

$$\frac{dx}{dt} = \omega(t, x). \tag{5.13}$$

The congruence of integral curves that it defines consists of what one calls *vortex lines* ([1]). Hence, the tangent to a vortex line at each point will point in the direction of the vorticity vector field at that point.

One naturally wonders how the vector field $\omega$ relates to the vector field $\mathbf{v}_s$. A good place to start is with their scalar product:

$$g(\mathbf{v}_s, \omega) = v_s(\omega) = v_s(\#^{-1}\omega) = v_s(\#^{-1} d_s v_s) = (v_s \wedge d_s v_s)(\mathbf{V}),$$

in which **V** is the 3-vector field that is dual to $V$ by way of $V(\mathbf{V}) = 1$. Since $V$ is globally non-zero, one can then say that:

**Theorem:** *For steady flow, the velocity vector field is orthogonal to the vorticity vector field iff the flow is completely integrable.*

A common construction that one encounters with vortex lines is a *vortex tube*, which one defines to be a 3-chain $c_3(0, 1)$ such that there are two cobounding 2-chains $c_2(0)$, $c_2(1)$ such that:

1. $\partial c_3(0, 1) = c_2(1) - c_2(0)$.
2. $\omega$ is interior normal to $c_2(0)$ and exterior normal to $c_2(1)$.

Hence, $c_3(0, 1)$ is simply a vector homology between the two 2-chains $c_2(0)$ and $c_2(1)$ with the vector field being $\omega$.

d. *Vorticity flux as a 2-coboundary.* The 2-form $\Omega$ can be integrated over 2-chains, and that integral will define a singular 2-cochain:

---

([1]) Which should not be confused with *line vortices*, which we will discuss below.

$$\Phi_{\boldsymbol{\omega}}[c_2] = \int_{c_2} \Omega. \qquad (5.14)$$

In the case of steady flow $\Omega = \omega$, and since $\omega = \#\boldsymbol{\omega}$, one can think of the 2-cochain $\Phi_{\boldsymbol{\omega}}$ as the *vorticity flux* through $c_2$, although the distinction will between steady and unsteady flow will complicate matters, but disappear for relativistic flow.

However, the fact that $\Omega = dv$ is also exact means that for any 2-cycle $z_2$ one will have:

$$\Phi_{\boldsymbol{\omega}}[z_2] = \int_{z_2} dv = \int_{\partial z_2} v = 0.$$

Hence, $\Phi_{\boldsymbol{\omega}}$ is a coboundary.

When two 2-chains $c_2$ and $c'_2$ are homologous (say, $c'_2 - c_2 = \partial c_3$), one will find that since $\Omega$ is closed, the vorticity flux will be the same for both of them:

$$\Phi_{\boldsymbol{\omega}}[c'_2 - c_2] = \Phi_{\boldsymbol{\omega}}[c'_2] - \Phi_{\boldsymbol{\omega}}[c_2] = \Phi_{\boldsymbol{\omega}}[\partial c_3] = \delta\Phi_{\boldsymbol{\omega}}[c_3] = 0.$$

Hence:

$$\Phi_{\boldsymbol{\omega}}[c'_2] = \Phi_{\boldsymbol{\omega}}[c_2].$$

We rephrase this as:

**Theorem:** *Vorticity flux is a homology invariant.*

Hence, the common value of $\Phi_{\boldsymbol{\omega}}[c_2]$ for homologous 2-chains will be referred to as the *strength* of the vortex tube $c_3$. Later on, we shall strengthen this theorem to vector homology.

*e. The magnetic analogy.* When Helmholtz first introduced the concept of vorticity [**40**], he also introduced an analogy between the concepts that were discussed in the context of vortical fluid flow and the ones that were becoming familiar in the theory of magnetism. In particular, he saw the velocity vector field as analogous to the **B** field of the latter theory and, for him, the vorticity vector field $\boldsymbol{\omega}$ played the role of the electric current that generated **B**. The solution of the basic equation $d_s v = \omega$ then took the form of a Biot-Savart construction – i.e., constructing the Green function for $d_s$. This analogy became especially appealing when one reduced the vortex tubes to filaments, which were then analogous to electric current-carrying wires.

However, that all happened before the days in which the theory of magnetism could be formulated in terms of differential forms. One finds that in three dimensions there are "accidental" linear isomorphisms between $\mathbb{R}^3$, its dual space, the space of bivectors over $\mathbb{R}^3$, and the space of 2-forms, since they are all three-dimensional. Hence, in the older vector calculus, it was not always clear whether one was referring to something that was actually a vector field, and not a covector field (i.e., 1-form), a bivector field, or a 2-form. Hence, the structural analogies between physical theories could be ambivalent in that sense.

For instance, if one regards **B** as the (spatial) dual of a spatial 2-form $B = \#\mathbf{B}$, which is closed ($d_s B = 0$), and thus, locally exact ($= d_s A$) then one could just as well make $B$ analogous to $\omega = d_s v$, which would make $A$ the analogue of $v$. This analogy is consistent with the Magnus force being the vortical analogue of the Lorentz force that a magnetic field exerts on an electric current, which we shall discuss below.

**6. Circulation as a 1-cochain.** – We have just seen that the vorticity flux $\Phi_\omega$ is a coboundary in two-dimensions. Hence, there will be a (non-unique) 1-cochain $c^1$ such that $\Phi_\omega = \delta c^1$. We shall now study the nature of any of the 1-cochains that are defined in that way.

*a. Basic definition.* By the definition of $c^1$, for any $c_2$:

$$\Phi_\omega[c_2] = \delta c^1[c_2] = c^1[\partial c_2] = \int_{\partial c_2} \alpha = \int_{c_2} d\alpha$$

for some 1-form $\alpha$.

However, we have already defined $\Phi_\omega$ by way of the 2-form $dv$, so we see that the 1-form $v$ can serve as the integrand for $c^1$. We then change its notation to something more conventional and define the 1-cochain:

$$C[c_1] = \int_{c_1} v \qquad (5.15)$$

to be the *circulation* of $c_1$.

When one takes the coboundary of $C$, one gets:

$$\delta C[c_2] = C[\partial c_2] = \int_{\partial c_1} v = \int_{c_1} dv = \Phi_\omega[c_2].$$

Hence:

**Theorem**: *Circulation will be a 1-cocycle iff the flow of $\mathbf{v}$ is irrotational.*

$v$ is exact – i.e., it admits a velocity potential $\psi$ – iff:

$$C[c_1] = \psi[\partial c_1] = \delta\psi[c_1],$$

so

$$C = \delta\psi,$$

and the circulation will be a coboundary.

In the event that $c_1$ is an elementary 1-simplex $\sigma_1$ and $\partial \sigma_1 = x_1 - x_0$, the 0-cochain that $\psi$ defines will be simply the *potential difference* cochain:

$$\Delta\psi[\sigma_1] \equiv \psi[\partial\sigma_1] = \psi(x_1) - \psi(x_0).$$

Thus, $\Delta\psi = \delta\psi$.

Typically, one considers the circulation around 1-cycles, and especially 1-boundaries, for which one will find that:

**Theorem:** *The following are equivalent:*

1. *Circulation is a 1-coboundary.*

2. *The circulation around every 1-cycle vanishes.*
3. *The covelocity admits a velocity potential.*
4. *v is exact.*

*The following are equivalent*:

1. *Circulation is a 1-cocycle.*
2. *The circulation around every 1-boundary vanishes.*
3. *The flow is irrotational.*
4. *v is closed.*

These statements include translations of basic properties of cocycles and coboundaries into the language of fluid kinematics.

*b. Topological sources of circulation for irrotational flows.* When a flow in *M* is irrotational, the circulation around every 1-boundary will vanish, but whether or not it also vanishes around every 1-cycle will depend upon the topology of *M*. In particular, for that to be true in every case, *M* would have to be simply-connected, which would make $H_1(M; \mathbb{R})$ and $H_{dR}^1(M)$ vanish.

If *M* is not simply-connected then every generator of $H_1(M; \mathbb{R})$ – i.e., every 1-cycle $z_1$ that does not bound – will be associated with non-zero circulation around it. One can think of the non-bounding 1-cycles in *M* as topological sources of circulation, even though they are associated with a flow of vanishing vorticity; we shall then refer to such a flow as having *topological vortices* in it.

One common topological source of "circulation-without-vorticity" is that of a *line vortex*, which basically amounts to the situation when the covelocity *v* is a basis element for $H_1(\mathbb{R}^3 - L; \mathbb{R}) \cong H_1(\mathbb{R}^2 - 0; \mathbb{R}) \cong H_1(S^1; \mathbb{R}) \cong \mathbb{R}$. Here, *L* is a line (or curve, for that matter) that is deleted from $\mathbb{R}^3$, such as the *z*-axis. Hence, as the first isomorphism shows, as far as homology is concerned, one could first make a deformation retraction of $\mathbb{R}^3 - L$ to $\mathbb{R}^2 - 0$, and then make a deformation retraction of the punctured plane to any circle in it that surrounds the missing point. A generator of $H_1(\mathbb{R}^3 - L; \mathbb{R})$ then amounts to a closed 1-form *v* that is not exact, such as $d\theta$ or, more generally, $f(\theta) d\theta$, where *f* is differentiable.

The *strength* of a line vortex is defined to be the circulation around any non-bounding 1-cycle in $\mathbb{R}^3 - L$, which will then be the same for all of them, since they all belong to the same homology class.

Another example of a topological source of circulation that is more topologically involved is when *v* is a basis element for $H_1(\mathbb{R}^3 - S^1; \mathbb{R})$, in which the space on which *v* is defined is now the complement of some circle $S^1$. That gives the example of a *ring* *vortex.* We shall not go into it further at the moment, but only point out the calculation of $H_1(\mathbb{R}^3 - S^1; \mathbb{R})$ involves "Alexander duality."

One of the key properties of irrotational flows is:

**Theorem:** *Circulation is a homology invariant for irrotational flows.*

The proof is straightforward: One chooses any two cobounding 1-chains $c_1$ and $c_1'$ that differ by $\partial c_2$ and evaluate the circulation of $c_1' - c_1$:

$$C[c_1' - c_1] = C[c_1'] - C[c_1] = C[\partial c_2] = \delta C[c_2] = \Phi_\omega[c_2] = 0.$$

Hence:

$$C[c_1'] = C[c_1].$$

**7. Basic fluid dynamics [19-22]** – We shall attempt to briefly summarize the key equations of fluid mechanics in a form that makes them amenable to discussions of a homological nature. In particular, one finds that homology first appears at the level of conservation laws, balance principles, and equilibrium.

*a. Homology and balance principles.* – One of the less-observed aspects of the conservation laws and balance principles that underlie the foundation of natural law is the fact that they can be given a natural homological context. This is exhibited most immediately when one looks at the flow of a scalar quantity through a region of space. Although the same argument works for any physical interpretation of the scalar quantity – e.g., mass, charge, particle number, etc. – we shall use mass for the sake of applications to fluid motions.

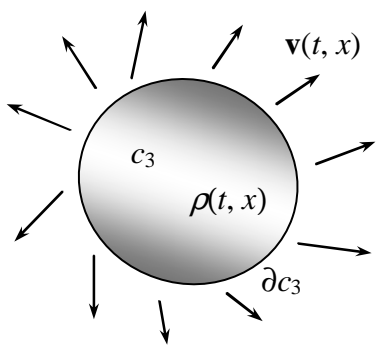

Figure 11. A 3-chain in a mass flow regime.

*b. Balance of mass.* Suppose that the mass in a space (an orientable differentiable manifold $\Sigma$ whose dimension needs only be greater than or equal to three) is distributed according to a mass density function $\rho(t, x)$ that is assumed to be smooth. Furthermore, it is in a state of motion with a velocity vector field $\mathbf{v}(t, x)$. One can then define a *mass current*:

$$\mathbf{p}(t, x) = \rho(t, x)\, \mathbf{v}(t, x), \tag{6.1}$$

which is seen to also coincide with the linear momentum density in this case.

Take any three-dimensional region of space that is represented by a 3-chain $c_3$ (e.g., a ball). We illustrate the situation so far in Fig. 11.

The *total mass* that is contained in $c_3$ is defined to be the integral of $\rho\, V = \#\rho$ over $c_3$:

$$M[c_3] = \int_{c_3} \#\rho\,. \tag{6.2}$$

Hence, one can regard total mass as a 3-cocyle; i.e., $M \in C^3(\Sigma; \mathbb{R})$.

As long as $\rho$ is a function of time, so is $M$. If we assume that $c_3$ and the volume element $V$ on $\Sigma$ are constant in time then we can define the *mass flow rate in $c_3$* to be the time derivative of $M$:

$$\dot{M}[c_3] = \frac{d}{dt}\int_{c_3} \#\rho = \int_{c_3} \#\frac{\partial \rho}{\partial t}\,. \tag{6.3}$$

One can also imagine that the mass is being transported (viz., convected) along the integral curves of $\mathbf{v}$ and define the *total mass flux* though a 2-chain $c_2$ to be the integral:

$$\Phi[c_2] = \int_{c_2} \#(\rho\mathbf{v})\,. \tag{6.4}$$

The 2-form $\#(\rho\, \mathbf{v})$ then takes on the interpretation of the mass flux density. Hence, we can regard $\Phi$ as a 2-cochain: $\Phi \in C^2(\Sigma; \mathbb{R})$.

In order to relate the mass flow rate to the total mass flux, one then poses the basic:

**Mass balance axiom:** *The mass flow rate in any 3-chain equals minus the total mass flux through its boundary:*

$$\dot{M}[c_3] = -\Phi[\partial c_3] = -\delta\Phi[c_3]. \tag{6.5}$$

Since this axiom is assumed to be true for any 3-chain, it will then lead to the fundamental:

**Theorem:** *The mass flow rate is a 3-coboundary, and in particular:*

$$\dot{M} = -\delta\Phi.$$

This has the:

**Corollary:** *The total mass in any 3-chain is conserved iff $\Phi$ is a 2-cocycle:*

$$\delta\Phi = 0. \tag{6.6}$$

By an application of Stokes's theorem to the integral form of the theorem, namely:

$$\int_{c_3} \# \frac{\partial \rho}{\partial t} = -\int_{\partial c_3} \#(\rho \mathbf{v}) \qquad \text{(for all } c_3\text{)}, \tag{6.7}$$

we can then derive the:

**Balance of mass equation (equation of continuity):**

$$0 = \frac{\partial \rho}{\partial t} + \text{div}(\rho \mathbf{v}). \tag{6.8}$$

One can define $x^0 = t$, $v^0 = 1$ and extend $\mathbf{v}$ to $\partial_t + \mathbf{v}$, so that this equation expresses merely the vanishing of a four-dimensional divergence:

$$0 = \frac{\partial (\rho v^\mu)}{\partial x^\mu} \qquad (\mu = 0, 1, 2, 3). \tag{6.9}$$

Another illuminating form that one can give to the equation of continuity is:

$$L_\mathbf{v} \rho = -\rho \, \text{div } \mathbf{v}, \tag{6.10}$$

i.e.:

$$\frac{1}{\rho} \frac{d\rho}{dt} = -\text{div } \mathbf{v}. \tag{6.11}$$

This allows one to state:

**Theorem:** *The density of a fluid is constant along the flow iff the fluid is incompressible.*

That does not, however, imply that the density must be constant in *space* for incompressible flow; the stronger condition on $\rho$ is only sufficient for incompressibility, but not necessary.

If $\Phi[.]$ is a non-cobounding 2-cocycle then the mass integral $M[.]$ will no longer be related to $\Phi[.]$.

*c. The balance of energy.* The balance of energy can also be given a homological expression by starting with a force 1-form $F$. The *total work* that is done along a 1-chain $c_1$ is defined to be:

$$W[c_1] = \int_{c_1} F. \tag{6.12}$$

Hence, the total work is a 1-chain: $W = C^1(\Sigma; \mathbb{R})$. We have:

**Theorem:** *The following are equivalent:*

1. *W is a 1-cocycle.*
2. *The work done around any 1-boundary vanishes.*
3. *F is closed.*
4. *The work done is a homology invariant.*

In the last statement, when one has $c_1' - c_1 = \partial c_2$ with $\partial c_1' = \partial c_1$, one will have:

$$W[c_1' - c_1] = W[c_1'] - W[c_1] = W[\partial c_2] = \delta W[c_2] = 0;$$

i.e.:

$$W[c_1'] = W[c_1].$$

We then define a *conservative* force by the:

**Theorem (conservation of work):** *The following are equivalent:*

1. *A force F is conservative.*
2. *The work done by the force is a 1-coboundary.*
3. *The work done by F around any 1-cycle vanishes.*
4. *$F = d\phi$ for some 0-form $\phi$.*
5. *The work done is a cobounding invariant.*

In the last statement, when one has $c_1' - c_1 = z_1$ with $\partial c_1' = \partial c_1$, and $W = \delta U$, one will have:

$$W[c_1' - c_1] = W[c_1'] - W[c_1] = \delta U[z_1] = U[\partial z_1] = 0;$$

i.e.:

$$W[c_1'] = W[c_1].$$

If the space $\Sigma$ is multiply-connected then it will be possible for there to be (non-bounding) 1-cycles around which the work is non-zero when it also vanishes around every 1-boundary.

*d. Balance of linear momentum.* – The equations of fluid motion derive from two balance principles, namely, the balance of mass and the balance of linear momentum. Since we have treated the former principle above, we shall now address the latter one.

We start by examining the nature of the Lie derivative of the *momentum density 1-form*:

$$p = \rho v. \tag{6.13}$$

along the flow of **v**. Hence, we shall examine:

$$L_\mathbf{v} p = (L_\mathbf{v} \rho) v + \rho L_\mathbf{v} v, \tag{6.14}$$

when we now employ the four-dimensional notations for **v** and $v$ (i.e., $\mathbf{v} = \partial_t + v^i \partial_i$, $v = dt + v_i\, dx^i$) in order to account for time-varying vector fields and mass densities.

If we assume balance of mass then, from (6.10), the first term in (6.14) can be replaced with $-\rho\, \text{div}\, \mathbf{v}$, and we will have:

$$L_{\mathbf{v}} p = \rho\, (a - \text{div}\, \mathbf{v}\, v). \tag{6.15}$$

in which $a = L_{\mathbf{v}} v$ is the convective acceleration 1-form that we discussed above.

This suggests the following:

**Theorem:** *The rate of change of linear momentum density along the flow of **v** is equal to $\rho\, a$ for all $t$ iff the flow is incompressible.*

We recall the expression (4.25) for $a$, and get:

$$L_{\mathbf{v}} p = \rho \left[ \frac{dv_s}{dt} + \tfrac{1}{2} dv^2 - (\text{div}\, \mathbf{v})\, v \right]. \tag{6.16}$$

We now introduce the *Euler equation* in its "infinite-dimensional $F = ma$" form:

$$\rho\, \frac{dv_s}{dt} = F - d_s \pi, \tag{6.17}$$

in which $F$ is the 1-form of external forces that act upon the fluid, and $\pi$ is the pressure in it.

Customarily, these equations are expressed in the equivalent form:

$$\frac{dv_s}{dt} = f - \frac{1}{\rho} d_s \pi. \tag{6.18}$$

in which $f = F / \rho$ is the force per unit mass (which is then an acceleration). The reason for that preference is probably due to the fact that many problems in fluid mechanics take the form of solving for $v$ when one is given the initial value of that vector field, and possibly its boundary values, if the fluid is confined to a channel or container.

If we extend $d_s$ to $d$ then we can write Euler's equation in the equivalent form:

$$L_{\mathbf{v}} p = F + d(\tfrac{1}{2}\rho v^2 - \pi) - \tfrac{1}{2} d\rho\, v^2 - \rho\, (\text{div}\, \mathbf{v})\, v. \tag{6.19}$$

The term in the first parentheses now represents the difference between the *dynamic pressure* $\tfrac{1}{2}\rho v^2$ (which is also the kinetic energy density) and the fluid pressure $\pi$.

Note that we have added an extra equation for the temporal component of $L_{\mathbf{v}}\, p$, namely:

$$0 = F_0 + \tfrac{1}{2} \rho\, \partial_t v^2 - \partial_t \pi, \tag{6.20}$$

which vanishes for steady flow iff $F_0 = 0$, such as when $F = -dU$, so $F_0 = -\partial_t U$. Here, we have used the fact that:

$$L_\mathbf{v}\, p = (L_\mathbf{v}\, \rho)\, dt + L_\mathbf{v}\, p_s = (-\rho \operatorname{div} \mathbf{v})\, dt + L_\mathbf{v}\, p_s.$$

*e. Balance of power.* Since $L_\mathbf{v}\, p = i_\mathbf{v}\, dp + d i_\mathbf{v}\, p$, we move the exact part of this expression [namely, $d(\rho v^2)$] to the right-hand side of (6.19) and get:

$$i_\mathbf{v}\, dp = F - d(\tfrac{1}{2}\rho v^2 + \pi) - \tfrac{1}{2} d\rho\, v^2 - \rho (\operatorname{div} \mathbf{v})\, v. \tag{6.21}$$

If we take the interior product of both sides with $\mathbf{v}$ one more time then the left-hand side must vanish, and we will get:

$$0 = i_\mathbf{v}\, F - L_\mathbf{v}(\tfrac{1}{2}\rho v^2 + \pi) + \tfrac{1}{2}(L_\mathbf{v}\rho)\, v^2 - \rho (\operatorname{div} \mathbf{v})\, v^2,$$

or, when we include the balance of mass:

$$P \equiv i_\mathbf{v}\, F = L_\mathbf{v}(\tfrac{1}{2}\rho v^2 + \pi) + \tfrac{1}{2}\rho (\operatorname{div} \mathbf{v})\, v^2. \tag{6.22}$$

This equation, which describes the balance of power, will be useful in the next section.

In particular, if $F = -dU$ then one can put (6.22) into the form:

$$\frac{dH}{dt} = -\tfrac{1}{2}\rho (\operatorname{div} \mathbf{v})\, v^2, \tag{6.23}$$

into which we have introduced the scalar function:

$$H = U + \tfrac{1}{2}\rho v^2 + \pi. \tag{6.24}$$

The function $H/g$ is referred to as the *total head* of the fluid.

Furthermore, since $dp = d\rho \wedge v + \rho \Omega_k$, which makes:

$$i_\mathbf{v}\, dp = L_\mathbf{v}\rho\, v - v^2\, d\rho + \rho\, i_\mathbf{v}\, \Omega_k,$$

we can also express (6.21) in the form:

$$i_\mathbf{p}\, \Omega = F - d(\tfrac{1}{2}\rho v^2 + \pi) + \tfrac{1}{2} d\rho\, v^2. \tag{6.25}$$

The right-hand side of this equation represents a force, and if one thinks of $\mathbf{p}$ as analogous to an electric current (when mass is replaced with charge) and $\Omega_k$ as analogous to the magnetic field strength 2-form $B$ (the covelocity is then replaced with the magnetic potential 1-form $A$) then the force $i_\mathbf{p}\Omega_k$ that would result would be analogous to the Lorentz force $i_\mathbf{J} B$ on a current $\mathbf{J}$ in a magnetic field. In the present context, one calls the force the *Magnus force*, and it actually explains such everyday phenomena as the deflection of a baseball when it is given a spin (e.g., curve balls and sliders), and even the lift on airplane wings.

When *F* is conservative, (6.25) can be put into the form:

$$i_{\mathbf{p}}\Omega = -dH + \tfrac{1}{2} d\rho\, v^2. \tag{6.26}$$

*f. Equation of state.* Between the equation of continuity and the three Euler equations, we have four first-order partial differential equations for five unknown functions, namely, $v^1$, $v^2$, $v^3$, $\rho$, $\pi$. In order to make the system well-determined, we need one more equation, which is usually given as an *equation of state* that expresses density as a smooth function of pressure:

$$\rho = \rho(\pi). \tag{6.27}$$

Of course, density might depend upon various other variables of a generally thermodynamics nature – especially temperature – but in the present case, one would call the fluid in question *barotropic*. A basic property of barotropic fluids is that the level surfaces of $\pi$ (viz., *isobars*) are parallel to the level surfaces of $\rho$ (viz., *isopycnics*), since:

$$d\rho \wedge d\pi = \frac{d\rho}{d\pi} d\pi \wedge d\pi = 0$$

*g. Steady flow.* If the fluid flow regime is steady then **v**, $v$, $\rho$, $\pi$, $F$ will all be functions of space alone, so:

$$\Omega = \omega = d_s v_s. \tag{6.28}$$

The equation of continuity will then reduce to:

$$0 = \operatorname{div}(\rho\, \mathbf{v}_s) = \operatorname{div} p_s, \tag{6.29}$$

or

$$\frac{1}{\rho} \mathbf{v}_s \rho = -\operatorname{div} \mathbf{v}_s. \tag{6.30}$$

This gives us a corollary to the theorem above:

**Corollary:** *For steady flow, if $\rho$ is constant in space (as well as time) then the flow will be incompressible.*

The Euler equation simplifies somewhat by the fact that all of the symbols in (6.19) are spatially-defined, including *d*. Furthermore, the path lines of **v** will coincide with its streamlines, and the convective derivative of $v_s$ will take the form:

$$\frac{dv_s}{dt} = (v^j \partial_j v_i)\, dx^j. \tag{6.31}$$

Now, assume that $F = -dU$ is conservative and that the fluid is incompressible. If we substitute that into the second of (6.23) then what will be left is:

$$L_v H = 0,$$

which gives:

**Bernoulli's theorem:** *The total head will be constant along the streamlines for an ideal, incompressible fluid in steady flow under the action of conservative external forces.*

When $U = $ const., one then sees the *Venturi effect* that the flow velocity is highest where the pressure is lowest, and vice versa.

**8. Dynamics of vortices.** – First, let us put Euler's equation into yet another convenient form:

$$L_v v = f + \tfrac{1}{2} dv^2 - \frac{1}{\rho} d\pi. \tag{6.32}$$

If the fluid is barotropic then:

$$d\left(\frac{1}{\rho} d\pi\right) = -\frac{1}{\rho^2} d\rho \wedge d\pi = -\frac{1}{\rho^2} \frac{d\rho}{d\pi} d\pi \wedge d\pi = 0,$$

so the final term in (6.32) will be a closed form. It can then be locally replaced with an exact one $d\vartheta$ (or globally if the manifold is simply-connected), and:

$$L_v v = f + d(\tfrac{1}{2} v^2 - \vartheta). \tag{6.33}$$

*a. Kelvin's circulation theorem* [**41**]. If we assume that $f$ is conservative then the right-hand side of (6.33) will be an exact 1-form; e.g., $d(\tfrac{1}{2} v^2 - \vartheta - u)$. Hence, $v$ will define a relative integral invariant. Of course, the integral of $v$ over a 1-cycle $z_1$ has already been defined to be the circulation of $v$ around $z_1$. Hence:

**Theorem:** *The circulation 1-cochain is a relative integral invariant for the inviscid flow of a barotropic fluid in a simply-connected flow region in the presence of conservative external volume forces.*
*It will be an absolute integral invariant iff $\tfrac{1}{2} v^2 - \vartheta - u$ is constant in time and space.*

One also finds that:

**Theorem:** *Circulation is a relative integral invariant for the flow of the vorticity vector field.*
*It will be an absolute integral invariant iff $v(\omega)$ is constant in time and space.*

This is true since:

$$L_\omega v = d(i_\omega v) + i_\omega dv = d(i_\omega v) + i_\omega \Omega = d(i_\omega v) = d(v(\omega)). \qquad (6.34)$$

(Here, we have used the fact that: $i_\omega \Omega = i_\omega \#\omega = i_\omega i_\omega V = 0$.) Note that this result is purely kinematical in character, so it is true independently of the fluid properties.

*b. Helmholtz's theorem* [**40**]. If one returns to equation (6.32) then one will note that if one takes the exterior derivative of both sides then one will get:

$$dL_v v = L_v dv = L_v \Omega = df. \qquad (6.35)$$

Hence:

**Helmholtz's theorem:** *Vorticity defines a relative integral invariant for the flow of any inviscid fluid.*
*It will be an absolute integral invariant iff the external forces are locally conservative.*

The first statement in this theorem is therefore true independently of whether the fluid is barotropic, the external force is conservative, or the flow is steady. Furthermore, since (1 / $\rho$) $d\pi$ does not have to be exact here, but only closed, the theorem will be true for multiply-connected flow regions.

One can make a stronger statement about the way that the vorticity varies along the vorticity vector field, since if we take the exterior derivative of both sides of (6.34) then we will get:

$$dL_\omega v = L_\omega dv = L_\omega \Omega = 0.$$

Hence:

**Theorem:** *Vorticity is always constant along the vortex lines.*

**9. Discussion.** – Now that we have seen how the methods of homology theory apply to the case of the non-relativistic motion of vortices, it will be simpler to go over the same notions in the context of relativistic fluid motion, which will be the second part of this monograph. In particular, the same techniques of homology will be used, so it will not be necessary to introduce more mathematics, only to revisit the basic fluid mechanics and the theory of vortices in particular. We will find that some things simplify, such as the fact that there is no distinction between time-varying and steady flow, so pathlines will coincide with streamlines. Other things will take on a different character that is mostly due to the expansion of the spatial dimension from three to four, although we have had a first exposure to that in the context of unsteady flow. The integrability of the covelocity and Poincaré duality will change noticeably, especially the definition of the vorticity vector field.

Some of the other topics in the theory of vortices that need to be addressed, besides its relativistic extension, are the role of viscosity, which is essential to the onset of turbulence, and some of the other singular distributions of vortices besides line vortices or ring vortices. For example, one also has vortex sheets.

Moreover, vortices are not the only topological defects in condensed matter physics that originate in non-trivial generators of the one-dimensional de Rham cohomology. When one goes to solid mechanics, one also encounters dislocations and disclinations, as well. However, one finds that these defects are also complicated by the fact that since elastic-plastic media admit strains, and more elaborate stresses than simple pressures, one will require more than one 1-form and one 2-form in order to describe matters, or rather differential forms with their values in something besides $\mathbb{R}$.

Of course, the very fact that the theory of vortices customarily requires an entire textbook to discuss thoroughly implies that one can expect to accomplish only so much within the scope of even a long article.

## References (¹)

---

(¹) References marked with an asterisk are available in English by D. H. Delphenich as free PDF downloads at neo-classical-physics.info.